\documentclass[journal=jctcce,manuscript=article,layout=twocolumn]{achemso}
\usepackage[colorlinks,linkcolor=blue,citecolor=blue,urlcolor=black,bookmarks=false,hypertexnames=true]{hyperref} 
\usepackage[utf8]{inputenc}
\usepackage[version=3]{mhchem}
\usepackage{amssymb}
\usepackage{color}
\usepackage{braket}
\usepackage{xspace}
\usepackage{graphicx}
\usepackage{subfig}
\usepackage{threeparttable}
\usepackage{multirow}
\usepackage{textcomp}
\usepackage{setspace}
\usepackage{caption}
\usepackage{array}
\usepackage{float}
\usepackage{cleveref}
\newcolumntype{L}[1]{>{\raggedright\let\newline\\\arraybackslash\hspace{0pt}}m{#1}}
\newcolumntype{C}[1]{>{\centering\let\newline\\\arraybackslash\hspace{0pt}}m{#1}}
\newcolumntype{R}[1]{>{\raggedleft\let\newline\\\arraybackslash\hspace{0pt}}m{#1}}
\allowdisplaybreaks
\makeatletter
\let\l@addto@macro\relax
\makeatother
\usepackage[fontsize=11pt]{scrextend}

\let\oldmaketitle\maketitle
\let\maketitle\relax
\renewcommand{\c}[1]{a^\dagger_{#1}}
\renewcommand{\a}[1]{a_{#1}}

\newcommand{\om}{\omega}
\newcommand{\si}{\sigma}
\newcommand{\ra}{\rightarrow}

\newcommand{\sub}[1]{\textsubscript{#1}}

\newcommand*{\maxe}{$\Delta_{\mathrm{MAX}}$\xspace}
\newcommand*{\mae}{\ensuremath{\Delta_{\mathrm{MAE}}}\xspace}
\newcommand*{\std}{\ensuremath{\Delta_{\mathrm{STD}}}\xspace}

\newcommand{\edge}[1]{\underline{\textrm{#1}}}

\crefname{figure}{Figure}{Figures}
\crefname{subfigure}{Figure}{Figures}
\crefname{table}{Table}{Tables}
\crefname{equation}{Eq.}{Eqs.}
\crefname{section}{Section}{Sections}
\crefname{subsection}{Section}{Sections}

\author{Ilia M.\ Mazin}
\email{mazin.3@osu.edu}
\affiliation{%
     Department of Chemistry and Biochemistry,
     The Ohio State University,
     Columbus, Ohio 43210, United States
}
 \author{Alexander Yu.\ Sokolov}
 \email{sokolov.8@osu.edu}
 \affiliation{%
     Department of Chemistry and Biochemistry,
     The Ohio State University,
     Columbus, Ohio 43210, United States
 }

\begin{tocentry}
\includegraphics{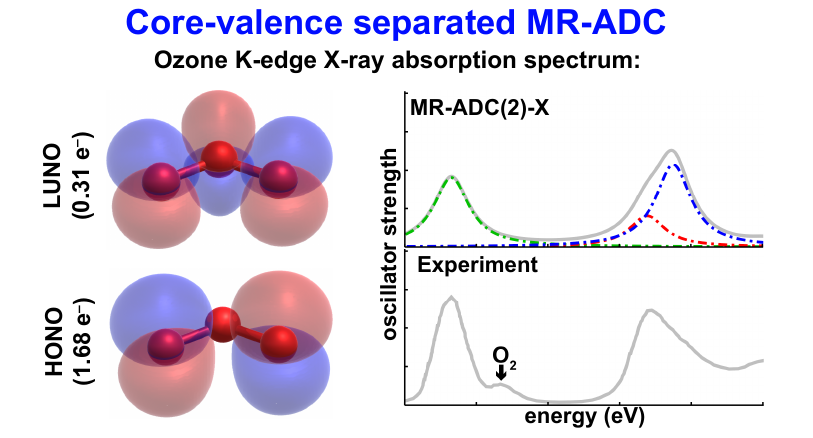}
\end{tocentry}

\title{{\color{blue}
Core-Excited States and 
X-Ray Absorption Spectra \\
From Multireference Algebraic Diagrammatic Construction Theory
}}
\begin{document}

%ArXiv %%%%%%%%%%%%%%
\newcommand*{\abstractext}{
We report the development and benchmark of multireference algebraic diagrammatic construction theory (MR-ADC) for the simulations of core-excited states and X-ray absorption spectra (XAS).
Our work features an implementation that incorporates core-valence separation into the strict and extended second-order MR-ADC approximations (MR-ADC(2) and MR-ADC(2)-X), providing an efficient access to high-energy excited states without including inner-shell orbitals in the active space.
Benchmark results on a set of small molecules indicate that at equilibrium geometries the accuracy of MR-ADC is similar to that of single-reference ADC theory when static correlation effects are not important.
In this case, MR-ADC(2)-X performs similarly to single- and multireference coupled cluster methods in reproducing the experimental XAS peak spacings.
We demonstrate the potential of MR-ADC for chemical systems with multiconfigurational electronic structure by calculating the K-edge XAS spectrum of the ozone molecule with a multireference character in its ground electronic state and the dissociation curve of core-excited molecular nitrogen.
For ozone, the MR-ADC results agree well with the data from experimental and previous multireference studies of ozone XAS, in contrast to the results of single-reference methods, which underestimate relative peak energies and intensities. 
The MR-ADC methods also predict the correct shape of core-excited nitrogen potential energy curve, in a good agreement with accurate calculations using driven similarity renormalization group approaches.
These findings suggest that MR-ADC(2) and MR-ADC(2)-X are promising methods for the XAS simulations of multireference systems and pave the way for their efficient computer implementation and applications. 
\vspace{0.25cm}
}
%ArXiv %%%%%%%%%%%%%%

%ArXiv %%%%%%%%%%%%%%
\twocolumn[
\begin{@twocolumnfalse}
\oldmaketitle
\vspace{-0.75cm}
\begin{abstract}
\abstractext
\end{abstract}
\end{@twocolumnfalse}
]
%ArXiv %%%%%%%%%%%%%%

\section{Introduction}
\label{sec:intro}

X-ray absorption spectroscopy (XAS) is a powerful element- and electronic-shell-specific experimental technique sensitive to the local chemical and electronic structure of the absorbing element.\cite{Wende:2004p2105,Stoehr:2013,Koide:2014p375} 
In XAS, X-ray light is used to excite the innermost core electrons into low-lying vacant orbitals, producing a spectrum that characterizes the system's local density of unoccupied states. 
Recent advances in X-ray radiation sources have enabled XAS experiments of a wide range of chemical systems with exceptional spatial and temporal resolution.\cite{Leone:2014p162,Nisoli:2017p10760,Kraus:2018p82,Bhattacherjee:2018p3203,Norman:2018p7208,Wernet:2019p20170464,Biswas:2022p893,Kobayashi:2022p0}.

\textit{Ab initio} electronic structure methods\cite{Szabo:1982,Helgaker:2000,Fetter:2003,Dickhoff:2008} play a crucial role in the analysis of experimental X-ray spectra that are often complex and can be difficult to interpret. 
Among the widely used theoretical methods for simulating XAS are 
time-dependent density functional theory (TD-DFT),\cite{Stener:2003p115,Nakata:2007p1295,Ekstrom:2006p143001,Song:2008p184113,Besley:2009p10350,Liang:2011p3540,Zhang:2012p194306,Lestrange:2015p2994,Besley:2016p5018,Wasowicz:2022p19302,Fenk:2022p26170,Konecny:2023p1360} 
equation-of-motion coupled cluster theory,\cite{Nooijen:1995p6735,Coriani:2015p181103,Peng:2015p4146,Vidal:2019p3117,Vidal:2020p2693,Folkestad:2020p6869,Nanda:2020p141104,Andersen:2022p6189,Petersen:2023p1775} 
configuration interaction,\cite{Roemelt:2013p204101,Maganas:2014p264,Maganas:2014p6374,Maganas:2018p1215,Oosterbaan:2018p044116,Wu:2018p245138,Kubas:2018p4320,Oosterbaan:2019p2966,Gerlach:2022p15217,Grofe:2022p10745} 
algebraic diagrammatic construction (ADC),\cite{Wenzel:2014p1900,Wenzel:2014p4583,Wenzel:2015p214104}
and related approaches.\cite{Schmitt:1992p1,Gilbert:2008p13164,Besley:2009p124308,Derricotte:2015p14360,Hait:2020p134108,Hait:2020p775,Garner:2020p154102,Peng:2019p1840}
A common approximation in all of these methods is that the ground-state electronic structure is well described using a single Slater determinant wavefunction (i.e., single-reference).

While this assumption is valid for many closed-shell molecules at (or near) equilibrium geometries, XAS is increasingly being used to probe core-level excited states of chemical systems with more challenging electronic structures, such as radicals,\cite{Alagia:2007p022509,Diaconescu:2007p3034,Fleet:2010p75,Sproules:2011p837,Sneeden:2017p11519,Ochmann:2017p4797,Nehzati:2018p10867,Schnorr:2019p1382,Wang:2019p323,Ephstein:2020p9524,Vidal:2020p9532,Kjellson:2020p236001} transition metal complexes,\cite{Groot:1989p5715,Hwang:2003p5791,Ray:2007p2783,Groot:2008p850,Mankad:2009p3878,Scarborough:2011p12446,Kurian:2012p452201,Wang:2013p4472,Oversteeg:2017p102,Baker:2017p182,Mortensen:2017p125136,Kubin:2018p6813,Lim:2019p6722,Lukens:2019p5044,Gaffney:2021p8010,Leahy:2022p9510} and along the dissociative reaction pathways\cite{Chen:2007p50,Zhou:2010p7727,Fabris:2011p5414,Stickrath:2013p4705,Tuxen:2013p2273,Xu:2016p4720,Morzan:2020p20044,Rott:2021p034104,Ledbetter:2022p9868,Ross:2022p9310} where multireference effects can be significant.
However, in contrast to single-reference theories, the range of available multireference methods is much narrower, including 
implementations of multiconfigurational self-consistent field,\cite{Agren:1989p745,Alagia:2013p1310,Engel:2014p1555,Guo:2016p3250,Pinjari:2016p477,Sergentu:2018p5583,Paris:2021p26559} multireference 
perturbation theories,\cite{Kunnus:2013p16512,Kunnus:2016p7182,Guo:2019p477,Nenov:2019p114110,Kallman:2020p8325,Northey:2020p2667,Segatta:2020p245,Koulentianos:2022p044306,Ghosh:2022p6323}
configuration interaction,\cite{Butscher:1977p449,Maganas:2014p20163,Coe:2015p58,Maganas:2019p104106,Jenkins:2022p141} coupled cluster,\cite{Brabec:2012p171101,Sen:2013p2625,Dutta:2014p3656,Maganas:2019p104106} 
and driven similarity renormalization group approaches.\cite{Huang:2022p219,Huang:2023p124112,Marin:2023p151101}

In this work, we present a new approach for simulating XAS of molecules with multiconfigurational electronic structure based on multireference algebraic diagrammatic construction theory (MR-ADC).\cite{Sokolov:2018p204113,Chatterjee:2019p5908,Chatterjee:2020p6343,Mazin:2021p6152,deMoura:2022p4769} 
The MR-ADC methods developed herein enable straightforward and computationally efficient simulations of core-excited states and XAS spectra by incorporating all single and double excitations from core molecular orbitals starting with a ground-state complete active-space wavefunction.
We describe the theory behind our implementation of MR-ADC in \cref{sec:theory}, followed by an overview of computational details (\cref{sec:computational_details}).
In \cref{sec:results}, we benchmark the MR-ADC methods for simulating core excitation energies and spectra in small weakly correlated molecules, the ozone molecule (\ce{O3}) with singlet diradical character in its ground state, and the dissociation curve of core-excited molecular nitrogen (\ce{N2}).
We outline our conclusions and goals for future developments in \cref{sec:conclusions}.

\section{Theory}
\label{sec:theory}

\subsection{Multireference Algebraic Diagrammatic Construction Theory for Electronic Excitations}
\label{sec:theory:mr_adc_ee}

In multireference algebraic diagrammatic construction theory (MR-ADC),\cite{Sokolov:2018p204113,Mazin:2021p6152,Chatterjee:2019p5908,Chatterjee:2020p6343,deMoura:2022p4769} electronically excited states are simulated by approximating the polarization propagator that describes how the electronic density of a chemical system is perturbed by an external electric field with frequency $\omega$.\cite{Fetter:2003,Dickhoff:2008,Schirmer:2008}. In its spectral form,\cite{Lehmann:1954p342} the polarization propagator ($\mathbf{\Pi} (\omega)$) can be written as
 \begin{align}
 	\label{eq:lehmann}
 	\Pi_{pqrs}(\om) &= \sum_{k\neq0}   \frac{\braket{\Psi | \c{q}\a{p} | \Psi_k} \braket{\Psi_k | \c{r} \a{s} | \Psi}}   {\om - (E_k - E)} \notag \\
 	                			&- \sum_{k\neq0}   \frac{\braket{\Psi | \c{r}\a{s} | \Psi_k} \braket{\Psi_k | \c{q} \a{p} | \Psi}}   {\om + (E_k - E)} 
 \end{align}
where it is expressed in terms of the fermionic creation ($\c{q}$) and annihilation ($\a{p}$) operators, the exact ground-state wavefunction $\ket{\Psi}$ with energy $E$, and the exact excited-state wavefunctions $\ket{\Psi_k}$ with energies ${E_k}$ ($k>0$).
As shown in \cref{eq:lehmann}, $\mathbf{\Pi} (\omega)$ contains the information about exact excitation energies ($E_k - E$) and transition probabilities ($\braket{\Psi | \c{q}\a{p} | \Psi_k} \braket{\Psi_k | \c{r} \a{s} | \Psi}$).
The two terms on the r.h.s.\@ of \cref{eq:lehmann}, known as the forward ($\mathbf{\Pi}_+(\omega)$) and backward ($\mathbf{\Pi}_-(\omega)$) components of the propagator, are related to each other as $\mathbf{\Pi}_+^\dag(-\omega) = \mathbf{\Pi}_-(\omega)$. 
For this reason, it is sufficient to focus on the forward component, which can be written in the tensor form as
\begin{align}
	\label{eq:lehmannmatrix}
	\mathbf{\Pi}(\omega) \equiv \mathbf{\Pi}_+(\omega) = \mathbf{\tilde{X}} (\om - \mathbf{\tilde{\Omega}})^{-1}  \mathbf{\tilde{X}}^\dag
\end{align}
where $\mathbf{\tilde{\Omega}}$ is a diagonal matrix of excitation energies ($\tilde{\omega}_k = E_k  - E$) and $\mathbf{\tilde{X}}$ is a tensor of spectroscopic amplitudes $\tilde{X}_{pqk} = \braket{\Psi | \c{q}\a{p} | \Psi_k}$. 

\cref{eq:lehmannmatrix} is valid only when written in the complete basis of electronic Hamiltonian eigenstates $\ket{\Psi_k}$.
Since the exact eigenstates are usually unknown, it is convenient to rewrite \cref{eq:lehmannmatrix} in a noneigenstate (but complete) basis of excited configurations\cite{Schirmer:1982p2395}
\begin{align}
	\label{eq:ADCprop}
	\mathbf{\Pi}(\omega) = \mathbf{T} (\omega \mathbf{S} - \mathbf{M})^{-1} \mathbf{T}^{\dag}
\end{align}
where $\mathbf{\tilde{\Omega}}$ is replaced by the nondiagonal effective Hamiltonian matrix $\mathbf{M}$, while $\mathbf{\tilde{X}}$ is substituted by the effective transition moments matrix $\mathbf{T}$.
For generality, the noneigenstate basis is assumed to be nonorthogonal with an overlap matrix $\mathbf{S}$.

\begin{figure}[t!]
	\includegraphics[width=0.4\textwidth]{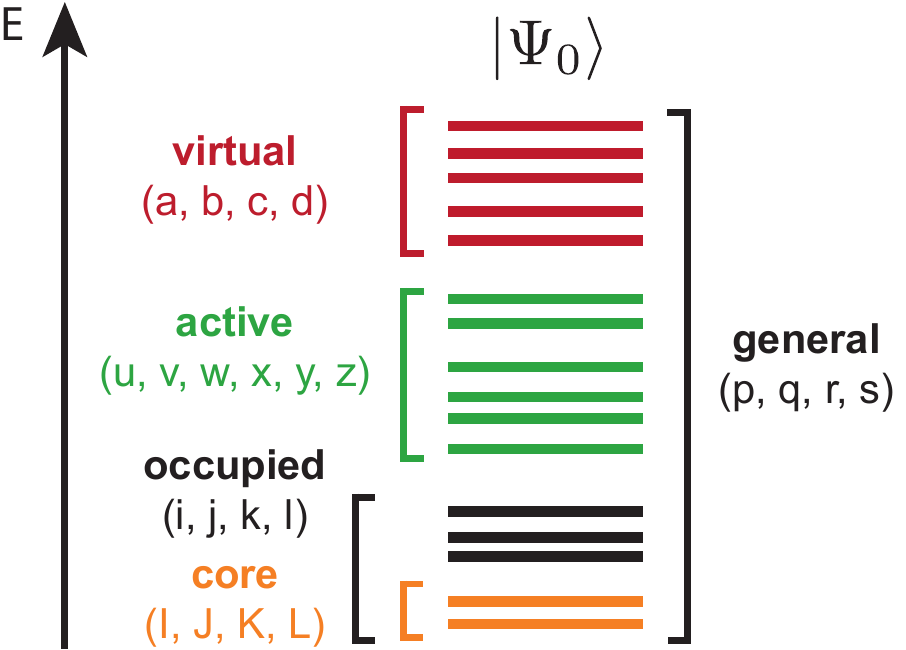}
	\captionsetup{justification=raggedright,singlelinecheck=false}
	\caption{Orbital spaces and their indices defined in this work.}
	\label{fig:cvs_mo_diagram}
\end{figure}

MR-ADC uses multireference perturbation theory to construct approximations for $\mathbf{\Pi}(\omega)$ in \cref{eq:ADCprop} and simulate properties of excited electronic states.\cite{Sokolov:2018p204113,Mazin:2021p6152}
First, molecular orbitals of the system are separated into occupied, active, and virtual subspaces (\cref{fig:cvs_mo_diagram}).
Next, the ground-state wavefunction $\ket{\Psi}$ is approximated by a multiconfigurational reference wavefunction $\ket{\Psi_0}$ obtained from a complete active-space self-consistent field (CASSCF) calculation.\cite{Knowles:1985p259,Werner:1985p5053,Malmqvist:1989p189} Following this, the electronic Hamiltonian $H$ is separated into the zeroth-order ($H^{(0)}$) and perturbation ($V = H - H^{(0)}$) contributions. 
Choosing $H^{(0)}$ as the Dyall Hamiltonian,\cite{Dyall:1995p4909,Angeli:2001p10252} each matrix in \cref{eq:ADCprop} is approximated up to the $n$th order in multireference perturbation theory
\begin{align}
	\label{eq:M_approx}
	\mathbf{M} &\approx \mathbf{M^{(0)}} + \mathbf{M^{(1)}} + \ldots + \mathbf{M^{(n)}} \\
	\label{eq:T_approx}
	\mathbf{T} &\approx \mathbf{T^{(0)}} + \mathbf{T^{(1)}} + \ldots + \mathbf{T^{(n)}} \\
	\label{eq:S_approx}
	\mathbf{S} &\approx \mathbf{S^{(0)}} + \mathbf{S^{(1)}} + \ldots + \mathbf{S^{(n)}}
\end{align}
defining the $n$th-order MR-ADC approximation (MR-ADC($n$)). 
Working equations for the matrix elements of $\mathbf{M}$, $\mathbf{T}$, and $\mathbf{S}$ at each order in perturbation are derived using the multireference formulation of effective Liouvillean theory.\cite{Prasad:1985p1287,Mukherjee:1989p257,Sokolov:2018p204113}

Solving the generalized eigenvalue problem 
\begin{align}
	\label{eq:eigenprob}
	&\mathbf{M} \mathbf{Y} = \mathbf{S} \mathbf{Y} \mathbf{\Omega}
\end{align}
allows to compute the MR-ADC($n$) excitation energies ($\mathbf{\Omega}$) and complementary eigenvectors ($\mathbf{Y}$), which can be used to calculate approximate spectroscopic amplitudes 
\begin{align}
	\label{eq:spec_amp}
	&\mathbf{X} = \mathbf{T} \mathbf{S}^{-\frac{1}{2}} \mathbf{Y}
\end{align}
Combining $\mathbf{X}$ with the matrix elements of dipole moment operator ($\mathbf{D}$) enables calculating oscillator strength
\begin{align}
	\label{eq:osc_strength}
	f_k = \frac{2}{3} \omega_k \left(\sum_{pq} D_{pq} X_{pqk}\right)^2
\end{align}
for each electronic transition with excitation energy $\omega_k$.
Furthermore, $\mathbf{\Omega}$ and $\mathbf{X}$ in \cref{eq:eigenprob,eq:spec_amp} can be used to define the MR-ADC($n$) polarization propagator
\begin{align}
	\label{eq:ADC_diag}
	\mathbf{\Pi}(\om) = \mathbf{X} (\om - \mathbf{\Omega})^{-1}  \mathbf{X}^\dag
\end{align}
which allows to simulate absorption spectra
by computing the spectral function 
\begin{equation}
	\label{eq:spectralfunction}
	A(\om) = -\frac{1}{\pi} \operatorname{Tr} \left[ \operatorname{Im} (\mathbf{D^{\dag}} \mathbf{\Pi}(\om) \mathbf{D}) \right]
\end{equation}

\begin{figure}[t!]
	\subfloat[]{\label{fig:M_matrix_strict}\includegraphics[width=0.35\textwidth]{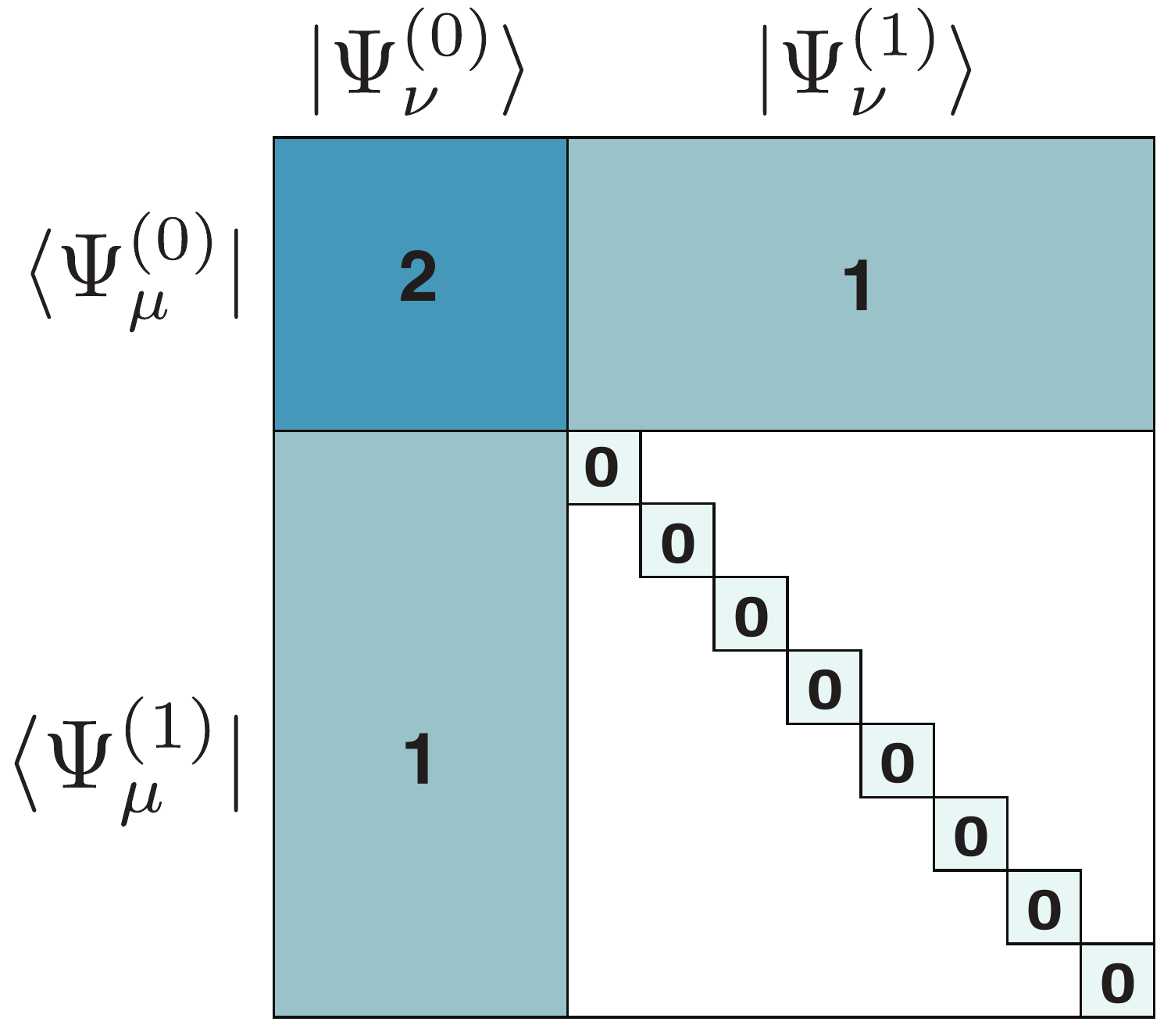}} \quad
	\subfloat[]{\label{fig:M_matrix_extended}\includegraphics[width=0.35\textwidth]{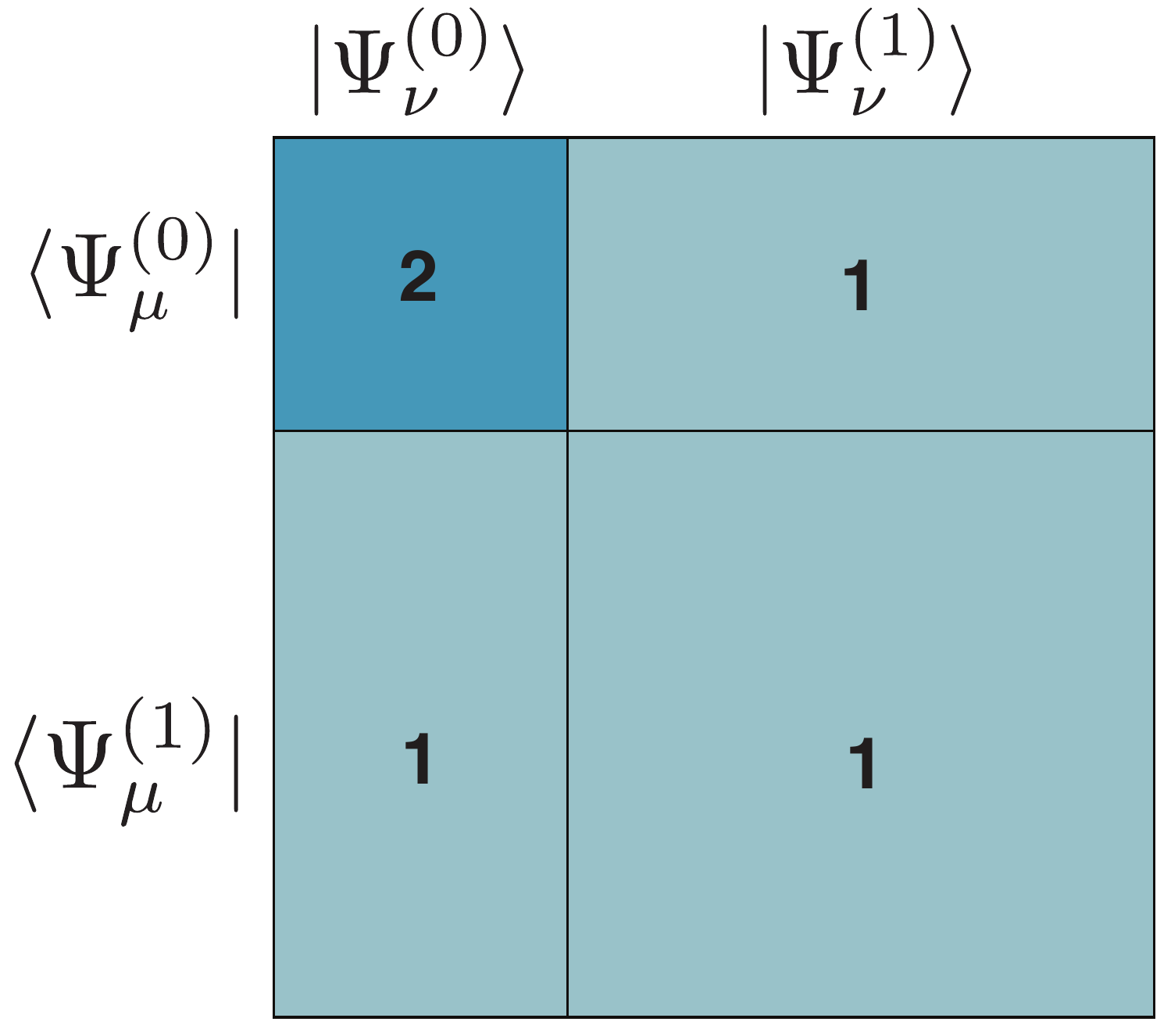}}
	\captionsetup{justification=raggedright,singlelinecheck=false}
	\caption{Perturbative structure of the effective Hamiltonian matrix $\mathbf{M}$ in (a) MR-ADC(2) and (b) MR-ADC(2)-X.
	Numbers denote the perturbation order to which the effective Hamiltonian is expanded for each sector. 
	Shaded areas indicate non-zero blocks.
	The excited electronic configurations $\ket{\Psi_\nu^{(l)}}$ $(l = 0,\ 1)$ are depicted in \cref{fig:full_manifold}.
}
	\label{fig:M_matrix}
\end{figure}

\begin{figure*}[t!]
	\includegraphics[width=0.90\textwidth]{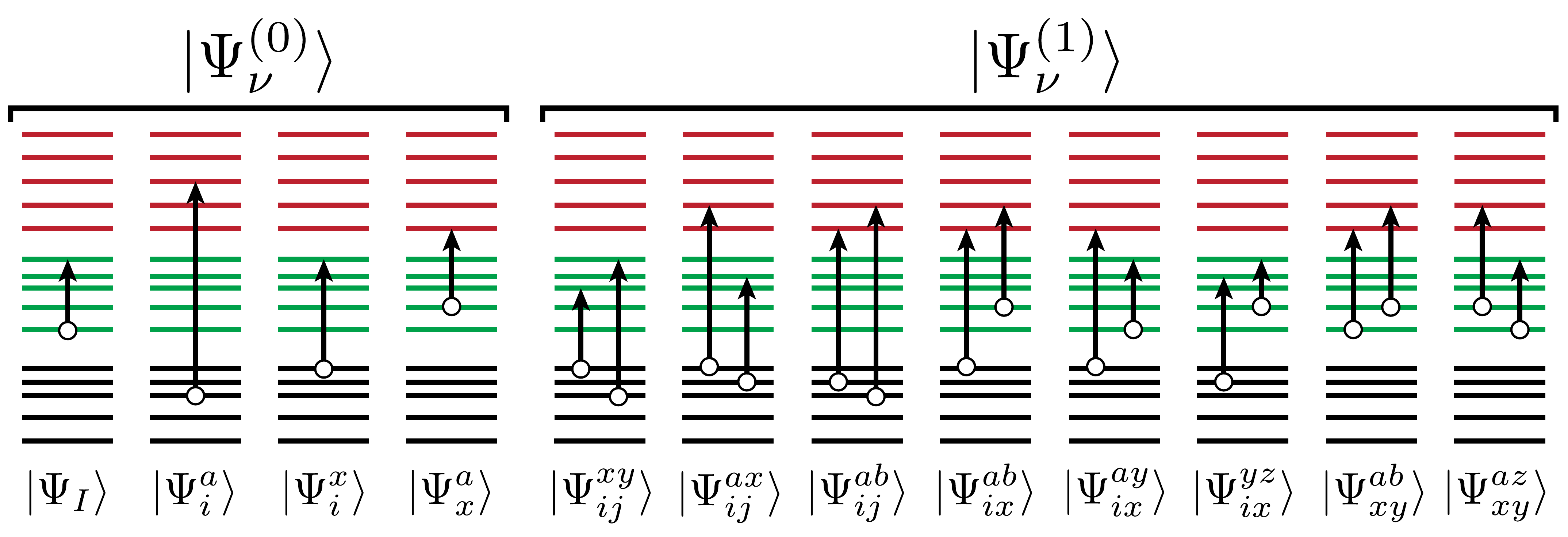}
	\captionsetup{justification=raggedright,singlelinecheck=false}
	\caption{Schematic illustration of the excited electronic configurations used to represent the elements of effective Hamiltonian matrix $\mathbf{M}$ in MR-ADC(2) and MR-ADC(2)-X (\cref{fig:M_matrix}). 
	A circle connected with an arrow denotes an excitation of one electron. 
	Black, green, and red lines represent occupied, active, and virtual orbitals, respectively (\cref{fig:cvs_mo_diagram}).
	}
	\label{fig:full_manifold}
\end{figure*}

\cref{fig:M_matrix} shows the perturbative structure of $\mathbf{M}$ matrix for the strict and extended second-order MR-ADC approximations: MR-ADC(2) and MR-ADC(2)-X, respectively.\cite{Mazin:2021p6152}
Each matrix element of $\mathbf{M}$ is evaluated with respect to a pair of excited configurations $\bra{\Psi_\mu^{(k)}}$ and $\ket{\Psi_\nu^{(l)}}$ where $k$ and $l$ indicate the perturbation order. 
Only the zeroth- ($\ket{\Psi_\nu^{(0)}}$) and first-order ($\ket{\Psi_\nu^{(1)}}$) excitations, schematically shown in \cref{fig:full_manifold}, appear in MR-ADC(2) and MR-ADC(2)-X.
The zeroth-order states $\ket{\Psi_\mu^{(0)}}$ describe all possible excitations in the active space ($\ket{\Psi_I}$) and single excitations between occupied, active, and virtual orbitals ($\ket{\Psi_i^a}$, $\ket{\Psi_i^x}$, and $\ket{\Psi_x^a}$).
The first-order configurations $\ket{\Psi_\mu^{(1)}}$ include all double excitations involving at least one non-active orbital and can be separated into eight excitation classes shown in \cref{fig:full_manifold}.
In each sector of $\mathbf{M}$ defined by $\bra{\Psi_\mu^{(k)}}$ and $\ket{\Psi_\nu^{(l)}}$, the MR-ADC(2) effective Hamiltonian is evaluated up to the order $m = 2 - k - l$.
MR-ADC(2)-X provides a higher-order treatment in the $\bra{\Psi_\mu^{(1)}}$ -- $\ket{\Psi_\nu^{(1)}}$ sector of $\mathbf{M}$, which improves the description of excited-state orbital relaxation effects and electronic states with double excitation character outside active space.

\subsection{Core-valence separated MR-ADC for simulating X-ray absorption}
\label{sec:theory:cvs_mr_adc}

In contrast to conventional multireference perturbation theories,\cite{Wolinksi:1989p3647,Hirao:1992p397,Werner:1996p645,Finley:1998p299,Andersson:1990p5483,Andersson:1992p1218,Angeli:2001p10252,Angeli:2001p297,Angeli:2004p4043} MR-ADC(2) and MR-ADC(2)-X incorporate the full spectrum of single and double excitations involving non-active molecular orbitals (\cref{fig:full_manifold}), such as the excited states of core electrons that are studied experimentally by X-ray absorption spectroscopy. 
However, the core-level excited states in MR-ADC calculations are deeply embedded in the eigenspectrum of effective Hamiltonian and are difficult to access without using special numerical techniques. 
Here, we employ the core-valence separation (CVS) approximation,\cite{Cederbaum:1980p206,Barth:1981p1038} which allows to efficiently compute core-excited states by neglecting their interaction with excitations from the remaining occupied orbitals. CVS has been widely used for simulating core-level excited states and X-ray absorption spectra in combination with a variety of electronic structure theories.\cite{Barth:1985p867,Trofimov:2000p483,Plekan:2008p360,Feyer:2009p5736,Wenzel:2014p1900,Coriani:2015p181103,Peng:2019p1840,Seidu:2019p144104,Vidal:2019p3117,Vidal:2020p2693,Fransson:2021p1618,Andersen:2022p6189,Petersen:2022p214305,deMoura:2022p4769,Seidu:2022p1345,Tenorio:2022p28150,Muchova:2023p6733,Datar:2023p1576,Petersen:2023p1775}

\begin{figure*}[t!]
	\includegraphics[width=0.90\textwidth]{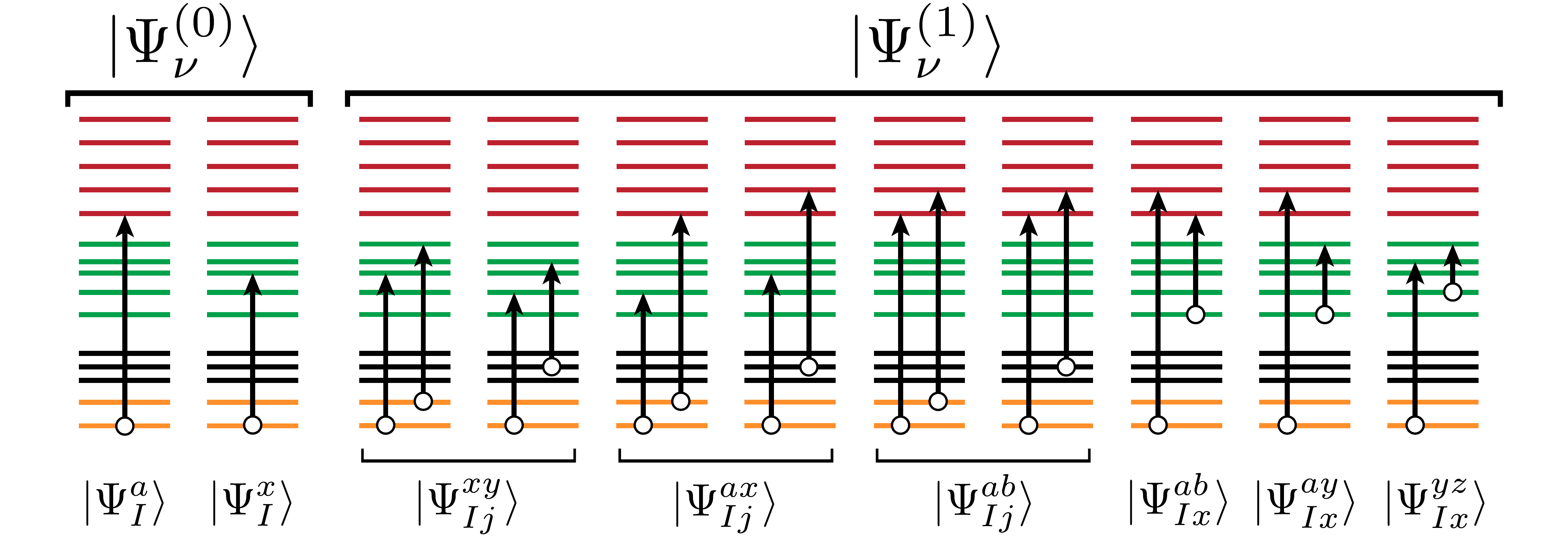}
	\captionsetup{justification=raggedright,singlelinecheck=false}
	\caption{
		Schematic illustration of the excited electronic configurations used to represent the elements of effective Hamiltonian matrix $\mathbf{M}$ in CVS-MR-ADC(2) and CVS-MR-ADC(2)-X (\cref{fig:M_matrix}). 
		A circle connected with an arrow denotes an excitation of one electron. 
		Orange, black, green, and red lines represent core, occupied, active, and virtual orbitals, respectively (\cref{fig:cvs_mo_diagram}).
	}
	\label{fig:cvs_manifold}
\end{figure*}

In our implementation of MR-ADC with CVS (CVS-MR-ADC), we define a new subspace of `core' molecular orbitals that includes the lowest-energy electrons, which are expected to participate in electronic transitions dominating the spectral region of interest (\cref{fig:cvs_mo_diagram}). 
Next, we introduce the CVS approximation by discarding the excitations without core indices from the MR-ADC configuration space (\cref{fig:full_manifold}).
The remaining configurations (\cref{fig:cvs_manifold}) are used to define the CVS-MR-ADC matrices ($\mathbf{M}$, $\mathbf{T}$, and $\mathbf{S}$), solve the generalized eigenvalue problem in \cref{eq:eigenprob} using the multiroot Davidson algorithm,\cite{Davidson:1975p87,Liu:1978p49} and compute excited-state properties and spectra (\cref{eq:spec_amp,eq:osc_strength,eq:ADC_diag,eq:spectralfunction}).
We note that the CVS approach used herein is different from the one employed in the CVS implementation of single-reference ADC by Wenzel and co-workers\cite{Wenzel:2014p1900,Wenzel:2014p4583,Wenzel:2015p214104} in two aspects: 
(i) it incorporates excitations with two core indices (\cref{fig:cvs_manifold}) while the implementation in Refs.\@ \citenum{Wenzel:2014p1900} -- \citenum{Wenzel:2015p214104} does not;
(ii) it does not employ frozen core approximation while the implementation in Refs.\@ \citenum{Wenzel:2014p1900} -- \citenum{Wenzel:2015p214104} does.

We verified our CVS-MR-ADC implementation against the results of MR-ADC calculations where CVS was introduced using the projector technique developed by Coriani and Koch.\cite{Coriani:2015p181103}
Reducing the configuration space in CVS-MR-ADC significantly lowers the computational cost relative to the full MR-ADC implementation.\cite{Mazin:2021p6152}
In particular, major computational savings are achieved by neglecting the all-active excitations (\cref{fig:full_manifold}) represented by the CASCI wavefunctions $\ket{\Psi_I}$ (\cref{sec:theory:mr_adc_ee}).
The resulting CVS-MR-ADC equations can be fully expressed and implemented in terms of only one type of reduced density matrices computed with respect to the reference CASSCF wavefunction.

\section{Computational Details}
\label{sec:computational_details}

The CVS-MR-ADC(2) and CVS-MR-ADC(2)-X methods were implemented in the development (spin-orbital) version of \textsc{Prism}, a Python-based implementation of MR-ADC methods.\cite{Prism:2023} The \textsc{Prism} program is interfaced with \textsc{PySCF}\cite{Sun:2020p024109} to obtain integrals and ground-state CASSCF wavefunctions. 
The MR-ADC calculations require specifying parameters to remove linearly dependent excited electronic configurations (\cref{fig:cvs_manifold}) and amplitudes of effective Hamiltonian.\cite{Chatterjee:2019p5908,Chatterjee:2020p6343}.
As in our previous work,\cite{Mazin:2021p6152} we used  $\eta_{d}$ = $10^{-10}$ for doubly external configurations and $\eta_{s}$ = $10^{-5}$ for single and semi-internal excitations.

In \cref{sec:results:small_molecules}, we benchmark the performance of CVS-MR-ADC for simulating K-edge core excitation energies in seven closed-shell molecules (\ce{C2H2}, \ce{C2H4}, \ce{CH4}, \ce{CO}, \ce{H2CO}, \ce{H2O}, and \ce{N2}). 
These calculations were performed at equilibrium geometries reported by Dutta et al.,\cite{Dutta:2014p3656}
with the exception of \ce{C2H4} where geometry was modified to restore the $D_{2h}$ point group symmetry (reported in the Supporting Information). 
The CVS-MR-ADC core excitation energies were compared to experimental data\cite{Hitchcock:2002p303,Schirmer:1993p1136,Hitchcock:1980p1,Puttner:1999p3415,Remmers:1992p3935,Myhre:2018p064106}, as well as the results from intermediate Hamiltonian Fock space multireference coupled cluster method (IH-FSMRCC)\cite{Dutta:2014p3656} and frozen-core CVS implementation of equation-of-motion coupled cluster theory with single and double excitations (\textit{fc}-CVS-EOM-CCSD).\cite{Vidal:2020p2693,Vidal:2019p3117,Coriani:2015p181103}
In addition, we performed calculations of core excitation energies using the CVS implementation of single-reference ADC (CVS-SR-ADC), which employed the same CVS scheme as the one used in CVS-MR-ADC (see \cref{sec:theory:cvs_mr_adc} for details).
The \textit{fc}-CVS-EOM-CCSD calculations were performed using \textsc{Q-Chem}\cite{qchem:50}, while the CVS-SR-ADC core excitation energies were computed using a development version of \textsc{PySCF}.\cite{Sun:2020p024109}
For all calculations in \cref{sec:results:small_molecules}, we used the aug-cc-pCVTZ basis set,\cite{Dunning:1989p1007,Kendall:1992p6796,Woon:1993p1358,Woon:1994p2975,Woon:1995p4572} with the exception of \ce{C2H2} and \ce{C2H4} where the \textit{f}-function with the smallest exponent was removed for each carbon atom to allow for a direct comparison with the IH-FSMRCC results reported in Ref.\@ \citenum{Dutta:2014p3656} where the same functions were omitted.
The details of active spaces used to perform the CVS-MR-ADC calculations are provided in the Supporting Information. 

In \cref{sec:results:ozone}, we apply CVS-MR-ADC along with CVS-SR-ADC and \textit{fc}-CVS-EOM-CCSD to simulate the X-ray absorption spectrum of ozone (\ce{O3}). 
Calculations were performed at the \ce{O3} equilibrium geometry from Ref.\@ \citenum{deMoura:2022p4769} (reported in the Supporting Information) and the aug-cc-pCVTZ-DK basis set.\cite{Woon:1995p4572,deJong:2001p48}
In addition, the exact two-component (X2C)\cite{Dyall:2001p9136,Liu:2009p031104} approach was used to account for scalar relativistic effects. 
The MR-ADC calculations were performed using the ground-state CASSCF reference wavefunction with 6 electrons in 8 active orbitals (6e, 8o) (\cref{fig:ozone_natocc}).
The ozone spectrum was simulated by computing and plotting a modified spectral function (\cref{eq:spectralfunction}) of the form:
\begin{equation}
	\label{eq:spectralfunction_plot}
	T(\om) = -\frac{1}{\pi} \operatorname{Im} \left[ \sum_{k} \frac{f_k}{\om - \om_k + i\mkern1mu \eta } \right]
\end{equation}
where $f_k$ and $\om_k$ are the oscillator strength (\cref{eq:osc_strength}) and energy of the $k$th transition, $\eta$ is a user-defined broadening parameter.

Finally, in \cref{sec:results:n2_pec}, we use the CVS-MR-ADC and CVS-SR-ADC methods to compute the potential energy curve of K-edge excited state along the bond dissociation coordinate in nitrogen molecule (\ce{N2}), which was recently studied by Huang and Evangelista using multireference driven similarity renormalization group approach (DSRG).\cite{Huang:2022p219} 
To compare the results of CVS-MR-ADC and DSRG methods directly, our study employs the cc-pCVQZ-DK basis set\cite{Dunning:1989p1007,Woon:1995p4572,deJong:2001p48} and X2C scalar relativistic Hamiltonian that were used in Ref.\@ \citenum{Huang:2022p219}. 
The CVS-MR-ADC calculations employed the (10e, 8o) full valence complete active space, which is equivalent to the GAS2 space used in the DSRG study.
The CVS-SR-ADC and CVS-MR-ADC total energies were computed by adding the CVS-SR-ADC and CVS-MR-ADC core excitation energies with the ground-state energies computed using second-order M\"oller--Plesset (MP2)\cite{Moller.1934} and partially-contracted N-electron valence (pc-NEVPT2)\cite{Angeli:2001p10252,Angeli:2001p297} perturbation theories, respectively. 
The MP2 and pc-NEVPT2 ground-state energies are reported in the Supporting Information.

For brevity, we will remove the `CVS' prefix in the abbreviations of CVS-MR-ADC, CVS-SR-ADC, and \textit{fc}-CVS-EOM-CCSD henceforth. 

\section{Results}
\label{sec:results}

\subsection{Core excitation energies of small molecules}
\label{sec:results:small_molecules}

\begin{figure}[t!]
	\subfloat[]{\label{fig:abs_error_plot}\includegraphics[width=0.48\textwidth]{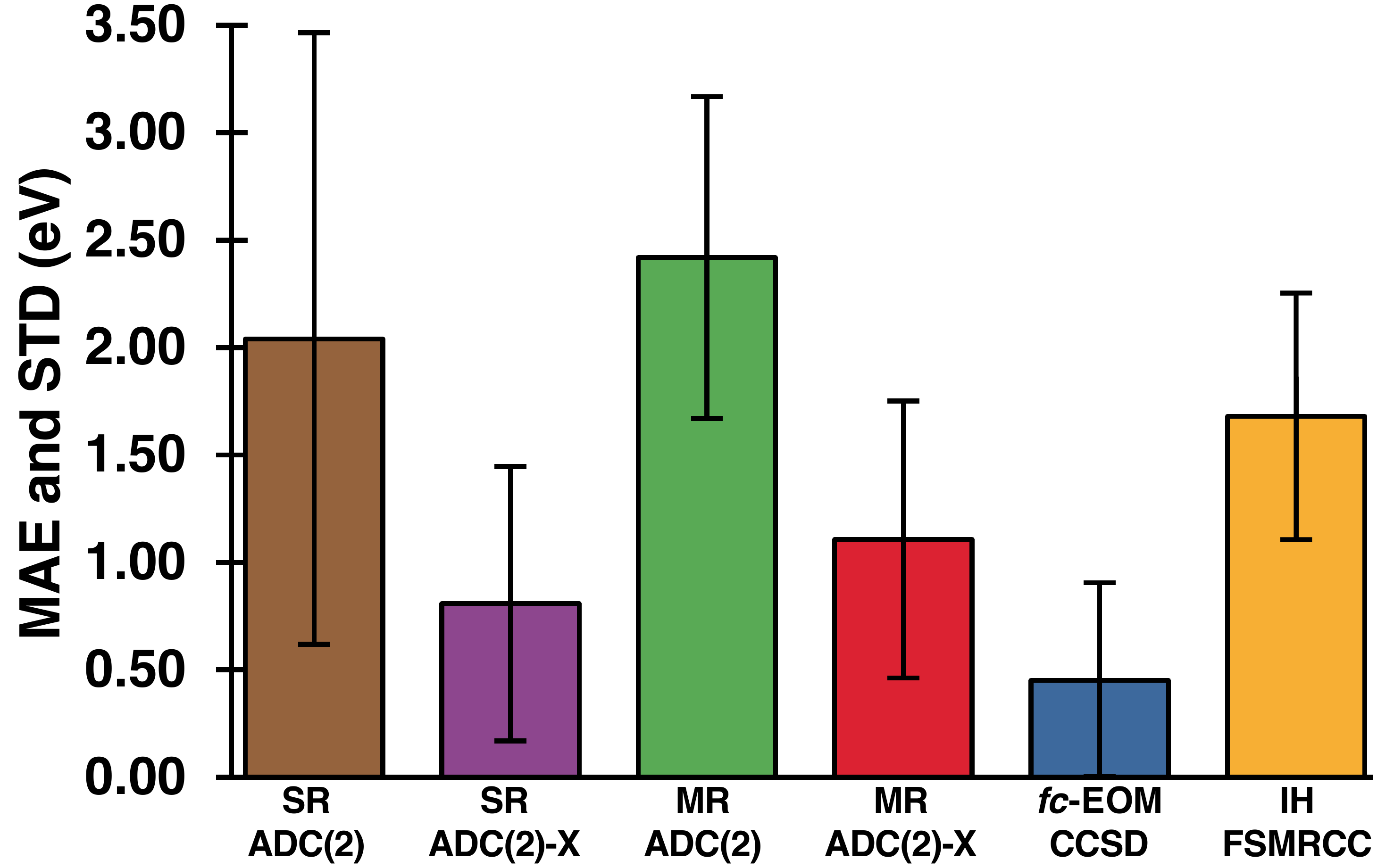}} \quad
	\subfloat[]{\label{fig:rel_error_plot}\includegraphics[width=0.48\textwidth]{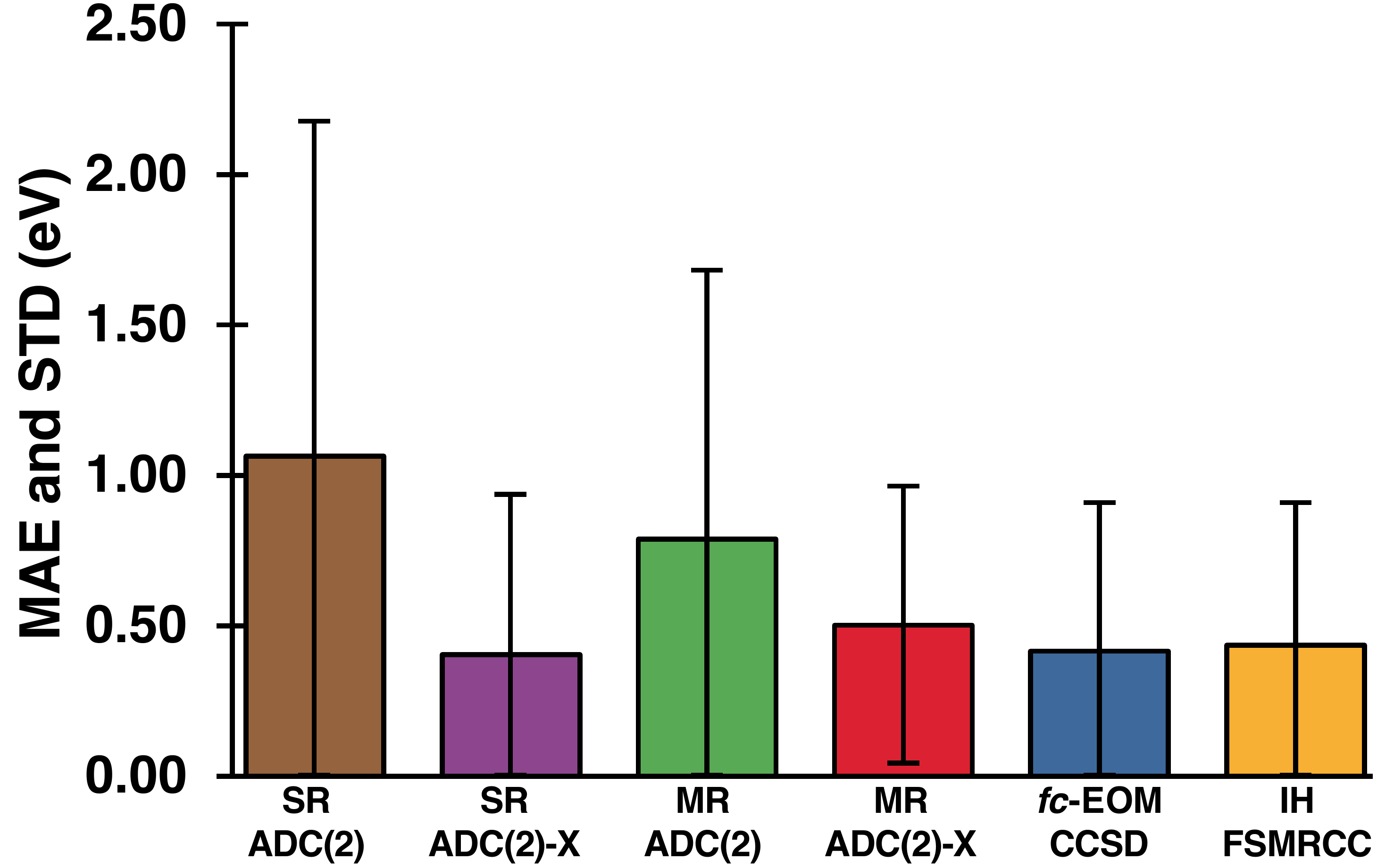}}
	\captionsetup{justification=raggedright,singlelinecheck=false}
	\caption{Mean absolute errors (\mae, eV) and standard deviations from the mean signed error (\std, eV) in core excitation energies (a) and peak spacings (b) computed using six methods relative to experimental results.\cite{Hitchcock:2002p303,Schirmer:1993p1136,Hitchcock:1980p1,Puttner:1999p3415,Remmers:1992p3935,Myhre:2018p064106}
		Each bar indicates \mae, the black vertical bracket denotes 2\std.
		See \cref{tab:ee_xray,tab:xray_errors} for details.
	}
	\label{fig:error_plots}
\end{figure}

We begin by benchmarking the performance of MR-ADC(2) and MR-ADC(2)-X for simulating vertical K-edge core excitation energies in seven closed-shell weakly correlated molecules at near-equilibrium geometries (\ce{C2H2}, \ce{C2H4}, \ce{CH4}, \ce{CO}, \ce{H2CO}, \ce{H2O}, and \ce{N2}).
\cref{tab:ee_xray} reports transition energies and oscillator strengths computed using MR-ADC, SR-ADC, \textit{fc}-EOM-CCSD, and IH-FSMRCC,\cite{Dutta:2014p3656} alongside with experimental results.\cite{Hitchcock:2002p303,Schirmer:1993p1136,Hitchcock:1980p1,Puttner:1999p3415,Remmers:1992p3935,Myhre:2018p064106}
We assess each method's accuracy by calculating mean absolute errors (\mae), standard deviations (\std), and maximum absolute errors (\maxe) in core excitation energies (CEE) and peak spacing (PS) relative to experiment.
These performance metrics are reported in \cref{tab:xray_errors} and are illustrated in \cref{fig:error_plots}.
We note that in our benchmark we focus on analyzing the performance of methods  {\it relative} to each other rather than assessing the ability of each method in reproducing the experimental results.
Since our simulations do not incorporate vibronic effects and calculate vertical excitation energies as opposed to the experimentally measured transition energies that are affected by molecular vibrations, direct comparison with the experimental results is not strictly valid.
%While we calculate the performance metrics of each method for both CEE and PS, we believe that the latter provides a better measure of a particular method's performance, since it is less likely to be affected by vibrational effects and differences in molecular geometry.

\begin{table*}[h!]
	\caption{Core excitation energies ($\Omega$, eV) and oscillator strengths ($f$) for the K-edge transitions of carbon, oxygen, and nitrogen atoms in seven small molecules computed using six methods, along with experimental results.\cite{Hitchcock:2002p303,Schirmer:1993p1136,Hitchcock:1980p1,Puttner:1999p3415,Remmers:1992p3935,Myhre:2018p064106}
		The element being probed is underscored. 
		See \cref{sec:computational_details} for the computational details.
		The IH-FSMRCC results are from Ref.\@ \citenum{Dutta:2014p3656}.
	}
	\label{tab:ee_xray}
	\setlength{\extrarowheight}{2pt}
	\setstretch{1}
	\tiny
	\centering
	\begin{threeparttable}
		\begin{tabular}{lccccccc}
			\hline\hline
			Excitation                       &   SR-ADC(2)    &  SR-ADC(2)-X   &   MR-ADC(2)    &  MR-ADC(2)-X   & \textit{fc}-EOM-CCSD & IH-FSMRCC & Experiment \\
			& $\Omega \ (f)$ & $\Omega \ (f)$ & $\Omega \ (f)$ & $\Omega \ (f)$ &    $\Omega \ (f)$    & $\Omega$  &  $\Omega$  \\ \hline
			\multicolumn{8}{c}{\textbf{System: \ce{\edge{C}2H2}}}                                                 \\
			$1s\ra\pi$*                      &    $288.7$     &    $285.1$     &    $287.9$     &    $284.7$     &       $285.7$        &  $287.1$  &  $285.7$   \\
			&   ($0.087$)    &   ($0.072$)    &   ($0.082$)    &   ($0.063$)    &      ($0.092$)       &           &            \\
			$1s\ra3s$                        &    $290.1$     &    $287.2$     &    $290.3$     &    $287.1$     &       $288.0$        &  $289.5$  &  $287.9$   \\
			&   ($0.002$)    &   ($0.002$)    &   ($0.007$)    &   ($0.002$)    &      ($0.006$)       &           &            \\
			$1s\ra3p$                        &    $290.5$     &    $287.6$     &    $290.9$     &    $287.4$     &       $289.3$        &  $289.8$  &  $288.9$   \\
			&   ($0.003$)    &   ($0.003$)    &   ($0.002$)    &   ($0.003$)    &      ($0.010$)       &           &            \\
			\multicolumn{8}{c}{\textbf{System: \ce{\edge{C}2H4}}}                                                 \\
			$1s\ra\pi_{b_{2g}}$              &    $287.9$     &    $284.3$     &    $287.0$     &    $283.5$     &       $284.9$        &  $285.8$  &  $284.7$   \\
			&   ($0.097$)    &   ($0.077$)    &   ($0.079$)    &   ($0.063$)    &      ($0.099$)       &           &            \\
			$1s\ra3s$                        &    $289.8$     &    $286.7$     &    $289.7$     &    $286.4$     &       $287.5$        &  $288.8$  &  $287.2$   \\
			&   ($0.007$)    &   ($0.006$)    &   ($0.008$)    &   ($0.006$)    &      ($0.009$)       &           &            \\
			$1s\ra3p$                        &    $290.5$     &    $287.4$     &    $290.6$     &    {$287.2$}     &       $288.2$        &  $289.3$  &  $287.9$   \\
			&   ($0.016$)    &   ($0.019$)    &   ($0.013$)    &   {($0.001$)}    &      ($0.029$)       &           &            \\
			\multicolumn{8}{c}{\textbf{System: \ce{\edge{C}H4}}}                                                 \\
			$1s\ra3p({t_2})$                 &    $290.3$     &    $287.6$     &    $290.7$     &    $287.6$     &       $288.0$        &  $289.4$  &  $288.0$   \\
			&   ($0.009$)    &   ($0.011$)    &   ($0.022$)    &   ($0.018$)    &      ($0.016$)       &           &            \\
			$1s\ra4p({t_2})$                 &    $291.8$     &    $288.7$     &    $292.1$     &    $288.7$     &       $289.2$        &  $291.1$  &  $289.5$   \\
			&   ($0.001$)    &   ($0.004$)    &   ($0.006$)    &   ($0.003$)    &      ($0.005$)       &           &            \\
			\multicolumn{8}{c}{\textbf{System: \ce{\edge{C}O}}}                                                  \\
			$1s\ra\pi$*                      &    $290.1$     &    $287.0$     &    $288.6$     &    $286.4$     &       $286.9$        &  $288.0$  &  $287.3$   \\
			&   ($0.083$)    &   ($0.066$)    &   ($0.057$)    &   ($0.046$)    &      ($0.081$)       &           &            \\
			$1s\ra3s$                        &    $296.0$     &    $292.7$     &    $295.7$     &    $292.1$     &       $292.9$        &  $294.2$  &  $292.5$   \\
			&   ($0.004$)    &   ($0.003$)    &   ($0.004$)    &   ($0.003$)    &      ($0.005$)       &           &            \\
			$1s\ra3p_\pi$                    &    $296.9$     &    $293.8$     &    $296.5$     &    $293.2$     &       $293.9$        &  $295.3$  &  $293.4$   \\
			&   ($0.006$)    &   ($0.008$)    &   ($0.015$)    &   ($0.017$)    &      ($0.007$)       &           &            \\
			$1s\ra4s$                        &    $297.1$     &    $294.1$     &    $296.7$     &    $293.5$     &       $294.1$        &  $295.3$  &  $294.8$   \\
			&   ($0.001$)    &   ($0.001$)    &   ($0.003$)    &   ($0.002$)    &      ($0.001$)       &           &            \\
			\multicolumn{8}{c}{\textbf{System: \ce{C\edge{O}}}}                                                  \\
			$1s\ra\pi$*                      &    $535.3$     &    $532.5$     &    $534.9$     &    $531.6$     &       $534.5$        &  $535.9$  &  $533.6$   \\
			&   ($0.033$)    &   ($0.032$)    &   ($0.033$)    &   ($0.028$)    &      ($0.041$)       &           &            \\
			$1s\ra3s$                        &    $538.5$     &    $537.2$     &    $541.0$     &    $537.3$     &       $539.4$        &  $540.8$  &  $538.9$   \\
			&   ($0.001$)    &   ($0.001$)    &   ($0.001$)    &   ($0.001$)    &      ($0.001$)       &           &            \\
			$1s\ra3p_\pi$                    &    $539.4$     &    $538.3$     &    $541.9$     &    $538.5$     &       $540.7$        &  $542.0$  &  $539.9$   \\
			&   ($<0.001$)   &   ($<0.001$)   &   ($<0.001$)   &   ($0.001$)    &      ($0.001$)       &           &            \\
			$1s\ra4s$                        &    $539.4$     &    $538.5$     &    $542.1$     &    $538.7$     &       $541.0$        &  $543.6$  &  $540.8$   \\
			&   ($<0.001$)   &   ($<0.001$)   &   ($<0.001$)   &   ($<0.001$)   &      ($<0.001$)      &           &            \\
			\multicolumn{8}{c}{\textbf{System: \ce{H2\edge{C}O}}}                                                 \\
			$1s\ra\pi$*                      &    $289.1$     &    $285.6$     &    $289.2$     &    $284.5$     &       $285.8$        &  $286.9$  &  $285.6$   \\
			&   ($0.067$)    &   ($0.051$)    &   ($0.053$)    &   ($0.042$)    &      ($0.065$)       &           &            \\
			$1s\ra3s$                        &    $293.5$     &    $290.2$     &    $293.5$     &    $289.7$     &       $290.5$        &  $291.8$  &  $290.2$   \\
			&   ($0.008$)    &   ($0.007$)    &   ($0.007$)    &   ($0.006$)    &      ($0.009$)       &           &            \\
			$1s\ra3p$                        &    $294.3$     &    $291.2$     &    $294.4$     &    $290.7$     &       $291.5$        &  $293.0$  &  $291.3$   \\
			&   ($0.020$)    &   ($0.019$)    &   ($0.019$)    &   ($0.018$)    &      ($0.026$)       &           &            \\
			\multicolumn{8}{c}{\textbf{System: \ce{H2C\edge{O}}}}                                                 \\
			$1s\ra\pi$*                      &    $532.7$     &    $529.7$     &    $531.5$     &    $528.6$     &       $531.3$        &  $533.1$  &  $530.8$   \\
			&   ($0.039$)    &   ($0.038$)    &   ($0.040$)    &   ($0.035$)    &      ($0.048$)       &           &            \\
			$1s\ra3s$                        &    $535.3$     &    $534.1$     &    $537.9$     &    $534.2$     &       $536.6$        &  $537.9$  &  $535.4$   \\
			&   ($0.001$)    &   ($0.001$)    &   ($0.001$)    &   ($0.001$)    &      ($0.001$)       &           &            \\
			$1s\ra3p$                        &    $536.2$     &    $535.0$     &    $539.4$     &    $535.2$     &       $537.6$        &  $538.9$  &  $536.3$   \\
			&   ($<0.001$)   &   ($0.001$)    &   ($<0.001$)   &   ($<0.001$)   &      ($<0.001$)      &           &            \\
			\multicolumn{8}{c}{\textbf{System: \ce{H2\edge{O}}}}                                                 \\
			$1s\ra3s(4a_1)$                  &    $534.7$     &    $532.8$     &    $536.4$     &    $532.0$     &       $534.4$        &  $535.9$  &  $534.0$   \\
			&   ($0.006$)    &   ($0.009$)    &   ($0.003$)    &   ($0.004$)    &      ($0.013$)       &           &            \\
			$1s\ra3p(2b_2)$                  &    $536.2$     &    $534.7$     &    $538.2$     &    $533.9$     &       $536.2$        &  $537.7$  &  $535.9$   \\
			&   ($0.009$)    &   ($0.018$)    &   ($0.008$)    &   ($0.014$)    &      ($0.026$)       &           &            \\
			$1s\ra3p(2b_1)$                  &    $538.2$     &    $536.5$     &    $539.9$     &    $535.5$     &       $538.1$        &  $539.4$  &  $537.0$   \\
			&   ($0.010$)    &   ($0.012$)    &   ($0.009$)    &   ($0.009$)    &      ($0.015$)       &           &            \\
			\multicolumn{8}{c}{\textbf{System: \ce{\edge{N}2}}}                                                  \\
			$\ce{N}_{\si_g}\ra1\pi_g(\pi^*)$ &    $403.6$     &    $399.9$     &    $402.8$     &    $400.6$     &       $400.8$        &  $401.9$  &  $400.9$   \\
			&   ($0.113$)    &   ($0.096$)    &   ($0.114$)    &   ($0.096$)    &      ($0.121$)       &           &            \\
			$\ce{N}_{\si_u}\ra4\si_g(3s)$    &    $408.3$     &    $405.6$     &    $410.1$     &    $406.4$     &       $406.8$        &  $407.6$  &  $406.2$   \\
			&   ($0.004$)    &   ($0.004$)    &   ($0.008$)    &   ($0.005$)    &      ($0.007$)       &           &            \\ \hline\hline
		\end{tabular}
	\end{threeparttable}
\end{table*}

\begin{table*}[t!]
	\caption{Mean absolute errors (\mae, eV), standard deviations (\std, eV), and maximum absolute errors (\maxe, eV) in core excitation energies (CEE) and peak spacings (PS) computed using six methods relative to experimental results.\cite{Hitchcock:2002p303,Schirmer:1993p1136,Hitchcock:1980p1,Puttner:1999p3415,Remmers:1992p3935,Myhre:2018p064106}
		The IH-FSMRCC results are from Ref.\@ \citenum{Dutta:2014p3656}.
		See \cref{tab:ee_xray} for data on individual molecules.
	}
	\label{tab:xray_errors}
	\setlength{\extrarowheight}{2pt}
	\setstretch{1}
	\scriptsize
	\centering
	\begin{threeparttable}
		\begin{tabular}{llcccccc}
			\hline\hline
			\multicolumn{2}{c}{}    & SR-ADC(2) & SR-ADC(2)-X & MR-ADC(2) & MR-ADC(2)-X & \textit{fc}-EOM-CCSD & IH-FSMRCC \\ \hline
			\multicolumn{8}{c}{\textbf{CEE:}}                                         \\
			\multicolumn{2}{c}{\mae}  &  $2.04$   &   $0.81$    &  $2.42$   &   $1.11$    &        $0.45$        &  $1.68$   \\
			\multicolumn{2}{c}{\std}  &  $1.42$   &   $0.64$    &  $0.75$   &   $0.64$    &        $0.46$        &  $0.58$   \\
			\multicolumn{2}{c}{\maxe} &  $3.56$   &   $2.30$    &  $3.97$   &   $2.27$    &        $1.29$        &  $2.85$   \\
			\multicolumn{8}{c}{\textbf{PS:}}                                         \\
			\multicolumn{2}{c}{\mae}  &  $1.06$   &   $0.40$    &  $0.79$   &   $0.50$    &        $0.42$        &  $0.44$   \\
			\multicolumn{2}{c}{\std}  &  $1.11$   &   $0.53$    &  $0.89$   &   $0.46$    &        $0.49$        &  $0.47$   \\
			\multicolumn{2}{c}{\maxe} &  $3.14$   &   $1.27$    &  $2.32$   &   $1.09$    &        $0.90$        &  $1.23$   \\ \hline\hline
		\end{tabular}
	\end{threeparttable}
\end{table*}

When considering CEE (\cref{fig:abs_error_plot}), the best agreement with experimental results is demonstrated by \textit{fc}-EOM-CCSD that shows \mae = $0.45$ eV and \std = $0.46$ eV.
Since the molecules in our benchmark set do not exhibit static electron correlation effects, the MR-ADC and SR-ADC methods show similar \mae at the same level of approximation.
The MR-ADC(2) method exhibits somewhat larger \mae  ($2.42$ eV) compared to that of SR-ADC(2) ($2.04$ eV), but has a much smaller \std (0.75 eV) that is by almost a factor of two smaller than \std of its single-reference counterpart (1.42 eV). 
Both methods show similar \maxe ranging between 3.56 to 3.97 eV.
We note that the computed errors in CEE originate from the difference of errors in the ground- and core-excited state energies.
For most of the electronic transitions in \cref{tab:ee_xray}, \textit{fc}-EOM-CCSD, SR-ADC(2), and MR-ADC(2) provide a better description of orbital relaxation and electron correlation effects in the ground electronic state than in the excited states, overestimating CEE relative to the experimental results.

Increasing the level of theory from ADC(2) to ADC(2)-X, reduces \mae in CEE for both MR- and SR-ADC by more than a factor of two. 
As a result, SR-ADC(2)-X and MR-ADC(2)-X show the best agreement with experiment among all ADC methods with \mae = $0.81$ and $1.11$ eV, respectively, and \std = 0.64 eV.
SR-ADC(2)-X and MR-ADC(2)-X also exhibit similar \maxe of $\sim$ 2.3 eV.
The reduction in CEE errors from ADC(2) to ADC(2)-X, which was also observed by Wenzel and co-workers in their benchmark of SR-ADC,\cite{Wenzel:2014p1900,Wenzel:2014p4583,Wenzel:2015p214104} can be attributed to an improved description of orbital relaxation effects provided by the extended ADC approximations that lowers the energies of core-excited states relative to the ground electronic state.
As a result, SR-ADC(2)-X and MR-ADC(2)-X tend to underestimate the experimental CEE for most electronic states in \cref{tab:ee_xray}.
The IH-FSMRCC method developed by Dutta et al.\cite{Dutta:2014p3656} shows intermediate performance between that of ADC(2) and ADC(2)-X methods with \mae = $1.68$ eV, \std = $0.58$ eV, and \maxe = $2.85$ eV in CEE.

The average errors of methods in predicting PS is illustrated in \cref{fig:rel_error_plot}.
MR-ADC(2)-X along with three other approaches (SR-ADC(2)-X, \textit{fc}-EOM-CCSD, and IH-FSMRCC) show similar performance with \mae ranging from 0.4 to 0.5 eV and \std $\sim$ 0.5 eV.
Larger differences are observed in \maxe, which increase in the following order: \textit{fc}-EOM-CCSD (0.90 eV) $<$ MR-ADC(2)-X (1.09 eV) $<$ IH-FSMRCC (1.23 eV) $<$ SR-ADC(2)-X (1.27 eV).
Increased errors in PS are shown by MR-ADC(2) and SR-ADC(2), which yield \mae and \std ranging from $0.79$ to $1.06$ eV and \maxe exceeding $2.3$ eV. 
Notably, all methods exhibit increased errors in PS for CO and \ce{H2CO}, suggesting that the description of electron correlation and orbital relaxation effects in the core-excited states of these two molecules is a challenge for linear-response and equation-of-motion theories employed in this work.
These two molecules also show the largest differences in ADC(2) and ADC(2)-X results indicating that the extended treatment of orbital relaxation in their core-excited states is important.

Overall, the performance of MR-ADC(2) and MR-ADC(2)-X for simulating CEE and PS in X-ray absorption spectra is consistent with the relative accuracy of SR-ADC for weakly correlated systems known from prior benchmarks.\cite{Wenzel:2014p1900,Wenzel:2014p4583,Wenzel:2015p214104} 
However, the MR-ADC methods are expected to be more reliable than the SR-ADC approximations when simulating core excitations in molecules with significant multireference character. 
We demonstrate this in \cref{sec:results:ozone} where we use MR-ADC to simulate the X-ray absorption spectrum of molecular ozone with multiconfigurational ground-state electronic structure and in \cref{sec:results:n2_pec} where we compute the potential energy curves for the K-edge excited state of nitrogen molecule.
%In \cref{sec:results:ozone}, we apply MR-ADC to simulate the X-ray absorption spectrum of molecular ozone with multiconfigurational ground-state electronic structure.

\subsection{X-ray absorption spectrum of ozone}
\label{sec:results:ozone}

\begin{figure*}[t!]
	\includegraphics[width=0.95\textwidth]{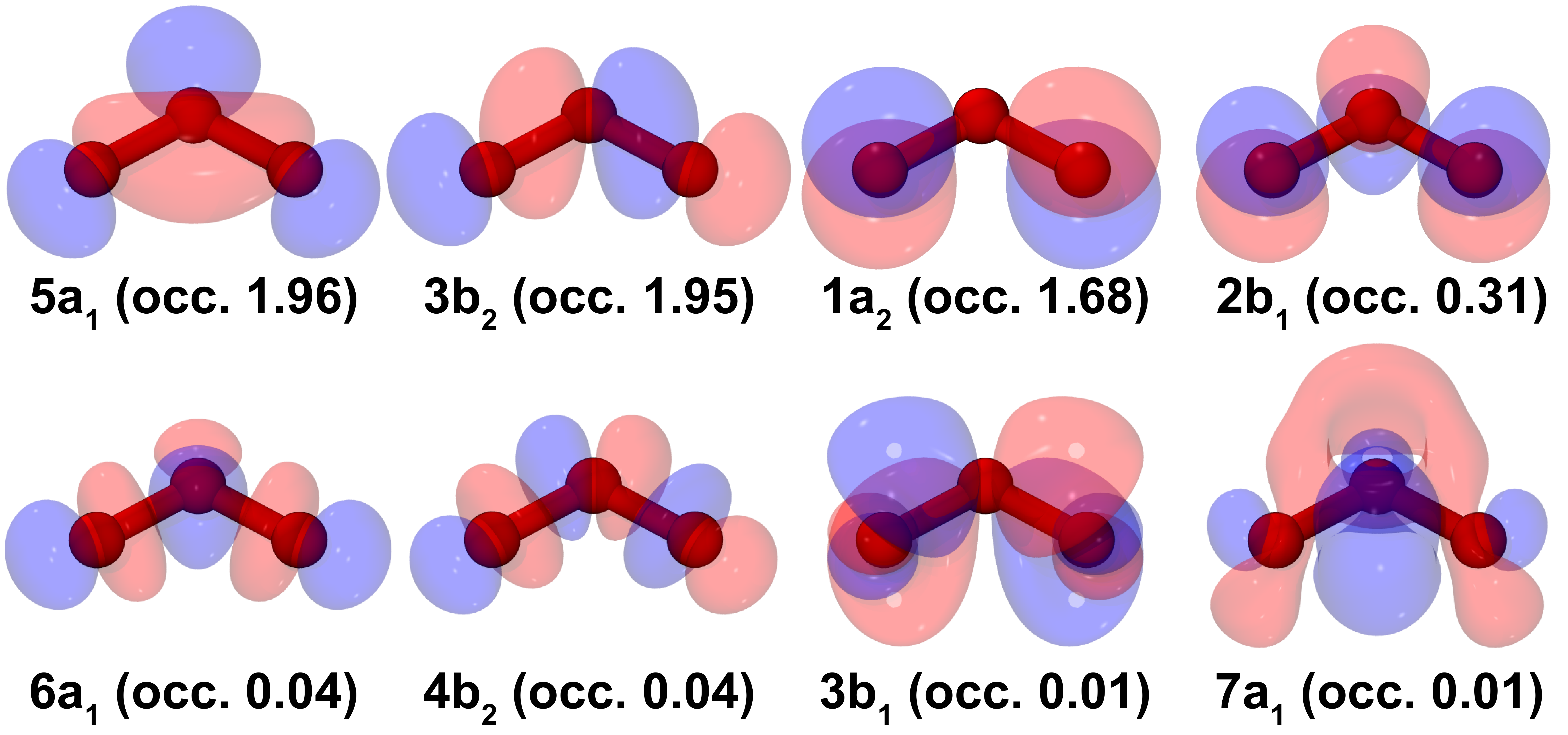}
	\captionsetup{justification=raggedright,singlelinecheck=false}
	\caption{Natural active orbitals and their occupation numbers for the ozone molecule computed using CASSCF with the (6e, 8o) active space and aug-cc-pCVTZ-DK basis set.
	}
	\label{fig:ozone_natocc}
\end{figure*}

\begin{figure*}[h!]
	\includegraphics[width=0.43\textwidth]{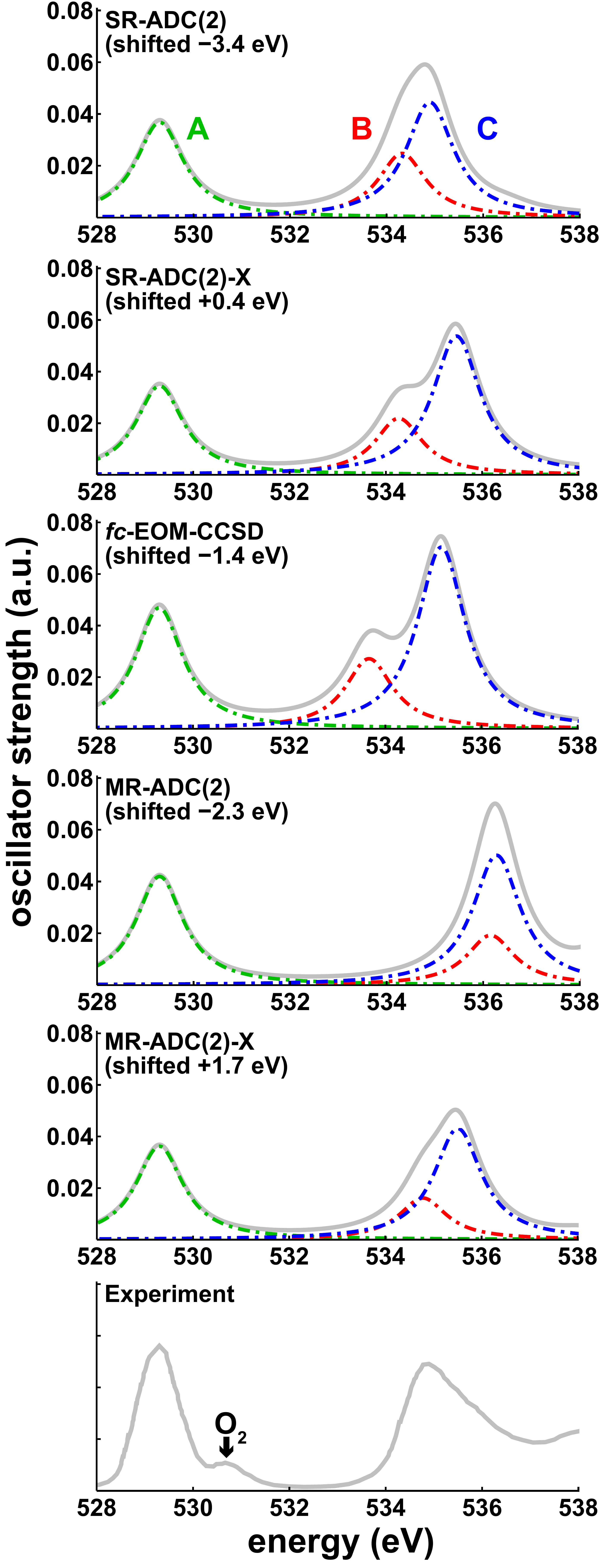}
	\captionsetup{justification=raggedright,singlelinecheck=false}
	\caption{Oxygen K-edge spectra of ozone molecule simulated using five methods in comparison to the experimental spectrum digitized\cite{Rohatgi:2022} from Ref.\@ \citenum{Stranges:2001p3400}.
		All theoretical spectra were plotted with $0.6$ eV broadening and were shifted to align the position of first peak with that in experimental spectrum. 
		The shift value is indicated on each plot. 
		Green, red, and blue lines correspond to the A, B, and C transitions in \cref{tab:ozone_peak_assign}, respectively.
		A small shoulder in the experimental spectrum is due to the presence of molecular oxygen.\cite{Stranges:2001p3400} 
	}
	\label{fig:ozone_kedge}
\end{figure*}

\begin{table*}[h!]
	\caption{Core excitation energies (eV) and oscillator strengths (in parentheses) for the three most prominent transitions (A, B, and C) in the oxygen K-edge X-ray absorption spectrum of ozone computed using several theoretical methods. 
	Also shown are A -- B and A -- C peak spacings ($\Delta_{\mathrm{BA}}$ and $\Delta_{\mathrm{CA}}$, eV).
	}
	\label{tab:ozone_peak_assign}
	\setlength{\extrarowheight}{2pt}
	\setstretch{1}
	\footnotesize
	\centering
	\begin{threeparttable}
		\begin{tabular}{lccccc}
			\hline\hline
			                              &                A                &                B                &                C                &                        &                        \\
			                              & \ce{O}\sub{T}$\ (1s) \ra \pi^*$ & \ce{O}\sub{C}$\ (1s) \ra \pi^*$ & \ce{O}\sub{T}$\ (1s) \ra \si^*$ &                        &                        \\
			Method                        &     ($2a_1/1b_2 \ra 2b_1$)      &        ($1a_1 \ra 2b_1$)        &     ($2a_1/1b_2 \ra 6a_1$)      & $\Delta_{\mathrm{BA}}$ & $\Delta_{\mathrm{CA}}$ \\ \hline
			SR-ADC(2)\tnote{a}            &             $532.7$             &             $537.7$             &             $538.2$             &         $5.0$          &         $5.5$          \\
			                              &            ($0.069$)            &            ($0.047$)            &            ($0.072$)            &                        &                        \\
			SR-ADC(2)-X\tnote{a}          &             $528.9$             &             $533.9$             &             $535.1$             &         $5.0$          &         $6.2$          \\
			                              &            ($0.065$)            &            ($0.041$)            &            ($0.082$)            &                        &                        \\
			\textit{fc}-EOM-CCSD\tnote{a} &             $530.7$             &             $535.0$             &             $536.6$             &         $4.3$          &         $5.9$          \\
			                              &            ($0.088$)            &            ($0.051$)            &            ($0.105$)            &                        &                        \\
			MC-RPA\tnote{b}               &             $538.0$             &             $543.6$             &             $544.8$             &         $5.7$          &         $6.8$          \\
			                              &            \tnote{f}            &            \tnote{f}            &            \tnote{f}            &                        &                        \\
			MS-RASPT2\tnote{c}            &             $529.0$             &             $535.8$             &             $536.4$             &         $6.8$          &         $7.5$          \\
			                              &            ($0.074$)            &            ($0.031$)            &            ($0.067$)            &                        &                        \\
			DSRG-MRPT2\tnote{d}           &             $530.7$             &             $535.9$             &             $537.1$             &         $5.2$          &         $6.4$          \\
			                              &            ($0.082$)            &            ($0.032$)            &            \tnote{f}            &                        &                        \\
			DSRG-MRPT3\tnote{d}           &             $528.4$             &             $533.5$             &             $535.0$             &         $5.1$          &         $6.6$          \\
			                              &            ($0.077$)            &            ($0.033$)            &            \tnote{f}            &                        &                        \\
			MR-LDSRG(2)\tnote{d}          &             $529.1$             &             $534.4$             &             $535.8$             &         $5.3$          &         $6.7$          \\
			                              &            \tnote{f}            &            \tnote{f}            &            \tnote{f}            &                        &                        \\
			MR-ADC(2)\tnote{a}            &             $531.6$             &             $538.5$             &             $538.6$             &         $6.9$          &         $7.0$          \\
			                              &            ($0.079$)            &            ($0.036$)            &            ($0.095$)            &                        &                        \\
			MR-ADC(2)-X\tnote{a}          &             $527.6$             &             $533.1$             &             $533.8$             &         $5.5$          &         $6.2$          \\
			                              &            ($0.069$)            &            ($0.030$)            &            ($0.081$)            &                        &                        \\
			{Experiment}\tnote{e}         &             {529.25}             &             {534.80}             &             {535.75}             &         {5.55}          &         {6.50}          \\ \hline\hline
		\end{tabular}
		\begin{tablenotes}\footnotesize
			\item[a] This work: (6e, 8o) active space for MR-ADC, aug-cc-pCVTZ-DK basis set.
			\item[b] From Ref.\citenum{Paris:2021p26559}: (12e, 12o) active space, aug-cc-pCVQZ basis set.
			\item[c] From Ref.\citenum{Tenorio:2022p28150}: (13e, 9o) in RAS2 and  (5e, 3o) in RAS1, cc-pVTZ basis set.
			\item[d] From Ref.\citenum{Huang:2023p124112}: (19e, 12o) in GAS2 and (5e, 3o) in GAS1, cc-pVQZ basis set.
			\item[e] From Ref.\citenum{Stranges:2001p3400}.
			\item[f] Oscillator strengths were not reported.
			\end{tablenotes}
	\end{threeparttable}
\end{table*}

Ozone (\ce{O3}) is a closed-shell molecule that has been a subject of many theoretical studies due to a significant multireference character in its ground electronic state.\cite{Hayes:1971p2090,Hay:1975p3912,Laidig:1981p3411,Kalemos:2008p054312,Miliordos:2013p5848,Miliordos:2014p2808,Takeshita:2015p7683} The multiconfigurational nature of \ce{O3} ground-state electronic structure can be observed by analyzing the CASSCF results reported in \cref{fig:ozone_natocc} where the occupations of frontier $1a_2$ and $2b_1$ natural orbitals deviate from two and zero by $\sim$ 0.3 $e^-$.
Several experimental\cite{Gejo:1997p497,Gejo:1999p4598,deBrito:2000p177,Stranges:2001p3400,Rosenqvist:2001p3614,Mocellin:2007p245,Frati:2020p4056,Tenorio:2022p28150} and theoretical\cite{Stranges:2001p3400,Vidal:2019p3117,Paris:2021p26559,Tenorio:2022p28150,Huang:2023p124112} studies of \ce{O3} core-excited states and X-ray absorption spectra (XAS) have been reported.
Here, we investigate the oxygen K-edge XAS spectrum of ozone using MR-ADC.

\cref{fig:ozone_kedge} compares the XAS spectra of \ce{O3} simulated using MR-ADC, SR-ADC, and \textit{fc}-EOM-CCSD to the experimental spectrum reported by Stranges et al.\cite{Stranges:2001p3400}. 
Two intense bands dominate the experimental K-edge spectrum: (i) a narrow feature at 529.3 eV and (ii) a broad peak centered at 534.9 eV.
The first band is attributed to the excitation from the terminal oxygen (\ce{O}\sub{T}) $1s$ orbitals to the $\pi^*$ ($2b_1$) frontier molecular orbital (transition A).
The second spectral feature is composed of overlapping peaks corresponding to two distinct electronic excitations: 
1) from the central oxygen (\ce{O}\sub{C}) $1s$ to the $\pi^*$ molecular orbital (transition B) 
and 2) from the terminal oxygen $1s$ to the $ \si^*$ ($6a_1$) molecular orbital (transition C).
Relative to A, the experimental excitation energies of B and C are $\Delta_{\mathrm{BA}} = 5.55$ eV and $\Delta_{\mathrm{CA}} = 6.50$ eV, respectively.\cite{Stranges:2001p3400}

All simulated XAS spectra in \cref{fig:ozone_kedge} correctly reproduce the ordering of A, B, and C transitions observed in the experiment, but differ in the first core excitation energy (A), peak spacings ($\Delta_{\mathrm{BA}}$ and $\Delta_{\mathrm{CA}}$), and relative intensities reported in \cref{tab:ozone_peak_assign}. 
Since our calculations do not incorporate vibronic effects, we focus on relative peak positions and intensities, which are expected to be less sensitive to changes in molecular geometry. 
The best agreement with the experimental spectrum is shown by MR-ADC(2)-X that underestimates $\Delta_{\mathrm{BA}}$ and $\Delta_{\mathrm{CA}}$ by 0.1 and 0.3 eV, respectively.
The SR-ADC and \textit{fc}-EOM-CCSD methods also underestimate peak spacings, but exhibit significantly larger errors ranging from 0.6 to 1.3 eV for $\Delta_{\mathrm{BA}}$ and from 0.3 to 1.0 eV for $\Delta_{\mathrm{CA}}$.
MR-ADC(2) is the only method that overestimates the experimental $\Delta_{\mathrm{BA}}$ and $\Delta_{\mathrm{CA}}$ showing errors of 1.3 and 0.5 eV, respectively.
The large differences in MR-ADC(2) and MR-ADC(2)-X results highlight the importance of differential orbital relaxation effects on the relative energies of \ce{O3} core-excited states.

\cref{tab:ozone_peak_assign} compares the MR-ADC results with the data from three recent multireference studies of \ce{O3} XAS: the calculation using multiconfigurational random phase approximation (MC-RPA) by Helmich--Paris,\cite{Paris:2021p26559} the work by Tenorio et al.\cite{Tenorio:2022p28150} using multistate restricted active space second-order perturbation theory (MS-RASPT2), and the study by Huang and Evangelista\cite{Huang:2023p124112} utilizing driven similarity renormalization group methods (DSRG).
We note that these three studies used different (sometimes non-core-polarized) basis sets and neglected scalar relativistic effects.
The MR-ADC(2) results compare well with the data from MS-RASPT2, which overestimates the experimental peak spacings by more than 1 eV.
The $\Delta_{\mathrm{BA}}$ and $\Delta_{\mathrm{CA}}$ computed using MR-ADC(2)-X are within $\sim$ 0.3 eV from the peak spacings calculated using the second- and third-order DSRG multireference perturbation theories (DSRG-MRPT2 and DSRG-MRPT3), which incorporate orbital relaxation effects in a state-specific fashion.
These results suggest that the MR-ADC methods are capable of achieving high accuracy in predicting XAS spectra of multireference systems without performing state-specific optimizations of excited-state wavefunctions.
Interestingly, the MC-RPA method, which is the lowest multireference level of theory listed in \cref{tab:ozone_peak_assign} bearing a close relationship with MR-ADC(1),\cite{Sokolov:2018p204113} shows errors in peak separations similar to MR-ADC(2)-X, which may be attributed to using a significantly larger active space and one-electron basis set. 

An attractive feature of MR-ADC methods is a straightforward access to oscillator strengths, which can be difficult to calculate using state-specific methods.\cite{Huang:2023p124112}
\cref{tab:ozone_peak_assign} compares the oscillator strengths ($f$) from MR-ADC and other theoretical approaches.
The \textit{fc}-EOM-CCSD method predicts the largest oscillator strengths for all transitions, possibly due to the frozen core approximation that neglects the dynamic correlation effects in polarization of core electron density upon excitation. 
The SR-ADC(2)-X and MR-ADC(2)-X oscillator strengths are very similar for A and C, but show a $\sim$ 27 \% decrease for B when multireference effects are included.
For all single-reference methods, the ratio of A and B oscillator strengths ($f$(A)/$f$(B)) is less than two, while $f$(A)/$f$(B) $>$ 2 is found in all multireference calculations.  
These differences in relative intensities of \ce{O}\sub{T}$\ (1s) \ra \pi^*$ and \ce{O}\sub{C}$\ (1s) \ra \pi^*$ transitions may be associated with the lack of static correlation description in the single-reference treatment of \ce{O}\sub{T} and \ce{O}\sub{C} $1s$ core-hole screening that has been recently discussed in Ref.\@ \citenum{deMoura:2022p4769}.

\subsection{Dissociation of core-excited nitrogen molecule}
\label{sec:results:n2_pec}

\begin{figure}[t!]
	\subfloat[]{\label{fig:n2_pec}\includegraphics[width=0.45\textwidth]{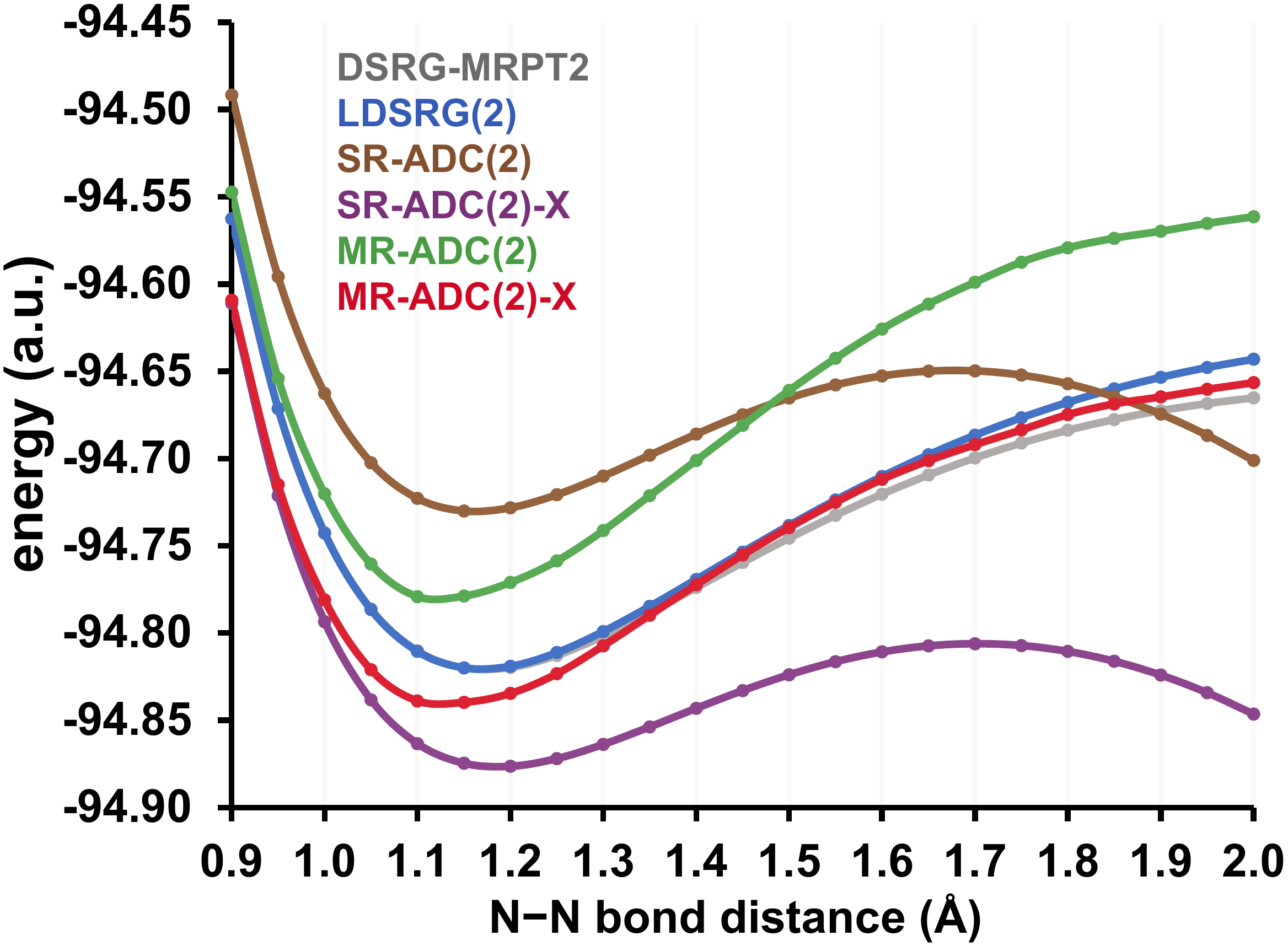}} \quad
	\subfloat[]{\label{fig:n2_error}\includegraphics[width=0.45\textwidth]{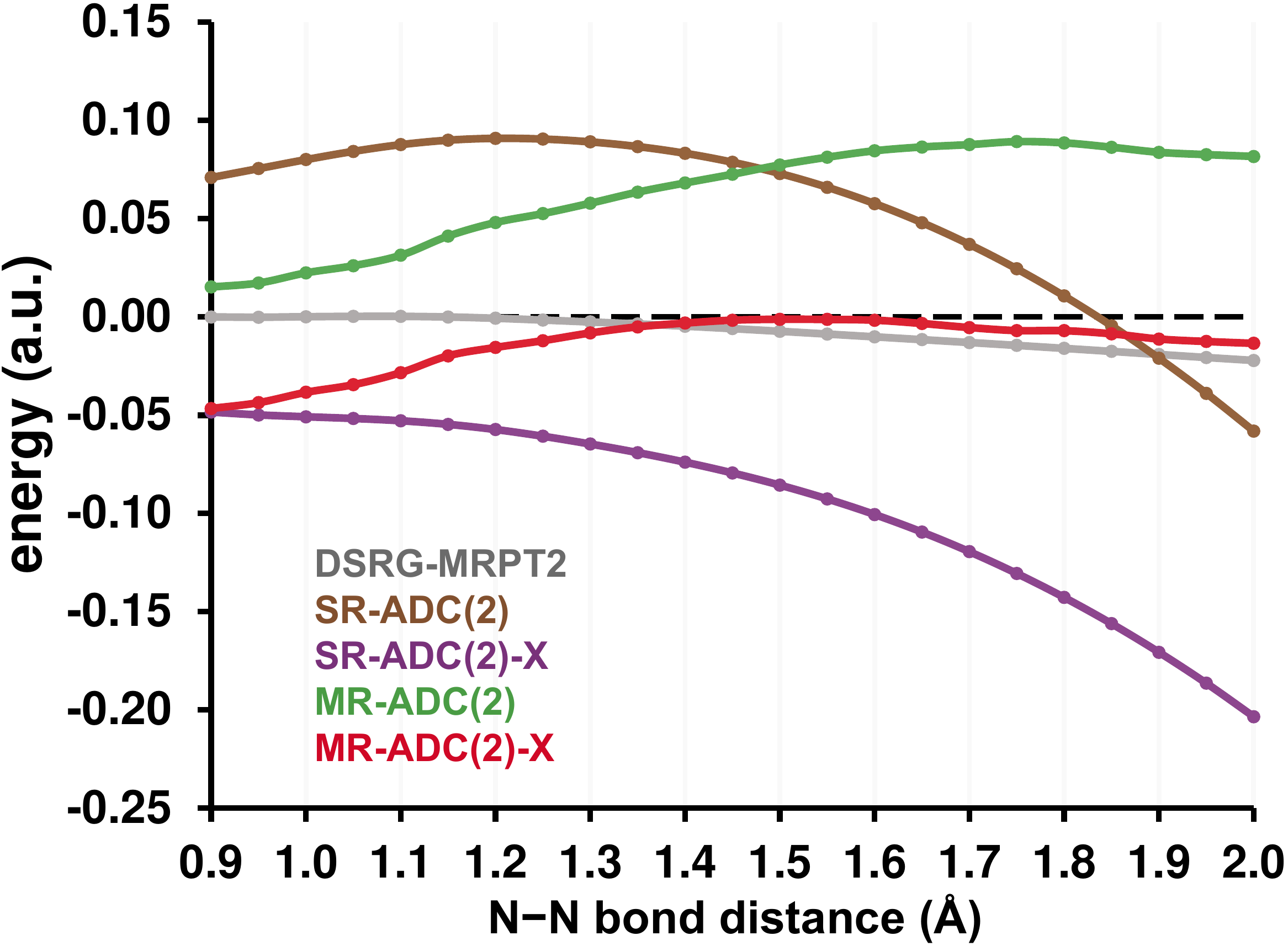}}
	\captionsetup{justification=raggedright,singlelinecheck=false}
	\caption{(a) Potential energy curves of molecular nitrogen (\ce{N2}) in the K-edge excited state computed using six different levels of theory with the cc-pCVQZ-DK basis set and X2C treatment of scalar relativistic effects. 
	(b) Error in the total energy of \ce{N2} K-edge excited state calculated using five methods relative to the LDSRG(2) potential energy curve.
	The DSRG results are from Ref.\@ \citenum{Huang:2022p219}.
	}
	\label{fig:n2_curve}
\end{figure}

Finally, to test the performance of MR-ADC away from equilibrium geometries, we compute the potential energy curve (PEC) of K-edge core-excited state in the \ce{N2} molecule that was recently studied by Huang and Evangelista using the DSRG-MRPT2 and MR-LDSRG(2) methods.\cite{Huang:2022p219}
We employ the same active space, basis set, and treatment of scalar relativistic effects as in Ref.\@ \citenum{Huang:2022p219}, which allows for a direct comparison of MR-ADC total energies with the PEC computed using the MR-LDSRG(2) method that incorporates dynamical correlation effects beyond second-order perturbation theory in a state-specific fashion.

\cref{fig:n2_pec} compares the \ce{N2} K-edge PEC computed using MR-ADC(2) and MR-ADC(2)-X with the potential energy curves simulated using SR-ADC(2), SR-ADC(2)-X, DSRG-MRPT2, and MR-LDSRG(2). 
In addition, in \cref{fig:n2_error}, we plot the error in total energy of the first five methods relative to MR-LDSRG(2).
Out of all ADC approximations, the best agreement with MR-LDSRG(2) is shown by MR-ADC(2)-X that overestimates the binding energy in the equilibrium region (for $r$(N--N) $<$ 1.3 \AA) but predicts energies close to DSRG-MRPT2 and MR-LDSRG(2) near dissociation ($r$(N--N) $>$ 1.3 \AA).
The MR-ADC(2) method underestimates the total energy of core-excited state across the entire PEC and exhibits a larger nonparallelity error compared to MR-ADC(2)-X, but predicts the correct shape of PEC in the dissociation region.
In contrast, the SR-ADC(2) and SR-ADC(2)-X methods produce unphysical PEC's that diverge away from the MR-LDSRG(2) curve near the dissociation limit with a barrier at $r$(N--N) $\approx$ 1.7 \AA.
These results suggest that the MR-ADC approximations are more reliable than the SR-ADC methods in the region of PEC where static correlation is important. 

\section{Conclusions}
\label{sec:conclusions}

In this work, we presented a new approach for simulating X-ray absorption spectra (XAS) of molecules using multireference algebraic diagrammatic construction theory (MR-ADC). 
Our work features an implementation of strict and extended second-order MR-ADC approximations (MR-ADC(2) and MR-ADC(2)-X) with core-valence separation (CVS) that enable direct calculations of core-excited states and spectra without including core orbitals in the active space.

To assess the accuracy of new MR-ADC methods, we first performed simulations of K-edge excitations in seven small molecules (\ce{C2H2}, \ce{C2H4}, \ce{CH4}, \ce{CO}, \ce{H2CO}, \ce{H2O}, \ce{N2}) at their equilibrium geometries. 
When compared to experimental results, MR-ADC(2)-X shows a significant improvement over MR-ADC(2) for core excitation energies and peak spacings due to a higher-level description of orbital relaxation effects. 
Our benchmark results indicate that for peak separations in XAS spectra MR-ADC(2)-X is competitive in accuracy with the frozen-core CVS implementation of equation-of-motion coupled cluster theory (\textit{fc}-EOM-CCSD) and a variant of Fock space multireference CC.\cite{Dutta:2014p3656}
Both MR-ADC(2) and MR-ADC(2)-X exhibit similar performance relative to their single-reference counterparts (SR-ADC(2) and SR-ADC(2)-X), which can be attributed to the lack of static correlation in the electronic structure of molecules in our benchmark set. 

Following our benchmark, we applied MR-ADC methods to simulate the oxygen K-edge XAS of the ozone molecule, which exhibits a multireference character in its ground electronic state. 
For this challenging system, the MR-ADC results are in a good agreement with the data from recent studies using multiconfigurational random phase approximation,\cite{Paris:2021p26559} restricted active space perturbation theory,\cite{Tenorio:2022p28150} and multireference driven similarity renormalization group (DSRG).\cite{Huang:2023p124112}
In contrast, single-reference methods (SR-ADC and \textit{fc}-EOM-CCSD) underestimate the spacing and relative intensities of $1s \ra \pi^*$ excitations originating from terminal and central oxygen atoms, likely missing contributions from static correlation in the description of core-hole screening effects.\cite{deMoura:2022p4769}
The MR-ADC(2)-X results show the best agreement with experiment predicting the peak spacings in XAS spectra with errors of 0.3 eV or less.

To further test how MR-ADC performs in multireference situations, we simulated the potential energy curve (PEC) of molecular nitrogen in the K-edge excited state using MR-ADC(2) and MR-ADC(2)-X and compared their results with accurate reference PEC computed using DSRG.\cite{Huang:2022p219}
Both MR-ADC(2) and MR-ADC(2)-X produce bound, qualitatively correct PEC with well-defined equilibrium and dissociation regions. 
The MR-ADC(2)-X method shows a good agreement with perturbative and non-perturbative variants of DSRG theory, while MR-ADC(2) is less quantitatively accurate. 
The SR-ADC methods are unable to correctly describe the K-edge PEC, showing an unphysical barrier near the dissociation region where static correlation effects are important.

The development of MR-ADC methods reported in this work opens up new opportunities for the simulations of X-ray absorption spectra in molecules with complicated electronic structure, such as transition metal complexes, open-shell molecules, and conjugated organic chromophores. 
To make these calculations possible, an efficient computer implementation of MR-ADC(2) and MR-ADC(2)-X needs to be developed, as opposed to a pilot (spin-orbital) code used in this work. 
Making this implementation applicable to a wide range of chemical systems would require extending MR-ADC to systems with open-shell ground states and incorporating spin-orbit coupling effects that are particularly important in XAS.
Work on these directions is currently ongoing in our group and will be reported in a separate publication.

\suppinfo
Information on composition of active spaces, ground-state CASSCF energies, molecular geometries of ethylene and ozone molecules, and ground-state potential energy curves of molecular nitrogen computed using RHF, MP2, CASSCF, and pc-NEVPT2 levels of theory.

\acknowledgement
This work was supported by the National Science Foundation under Grant No.\@ CHE-2044648. 
Computations were performed at the Ohio Supercomputer Center under the project PAS1583.\cite{OhioSupercomputerCenter1987}
The authors would like to thank Meng Huang and Francesco Evangelista for providing the DSRG potential energy curves for the core-excited state of nitrogen molecule.

%\bibliography{ilia_papers,refs_alex}

\providecommand{\latin}[1]{#1}
\makeatletter
\providecommand{\doi}
  {\begingroup\let\do\@makeother\dospecials
  \catcode`\{=1 \catcode`\}=2 \doi@aux}
\providecommand{\doi@aux}[1]{\endgroup\texttt{#1}}
\makeatother
\providecommand*\mcitethebibliography{\thebibliography}
\csname @ifundefined\endcsname{endmcitethebibliography}
  {\let\endmcitethebibliography\endthebibliography}{}

\end{document}